\documentclass[aps,prl,showpacs,amsmath,twocolumn]{revtex4}
\usepackage{graphicx
}
\usepackage{epsf}
\usepackage{subfigure}
\begin{document}
\title{Hysteresis of
Finite Arrays of Magnetic Nano Dots.}
\author{M. Amin Kayali and Wayne M.
Saslow}
\affiliation{Department of Physics, Texas A \& M University,
College Station, 
Texas 77843-4242, USA.}
\begin{abstract}
Hysteresis curves for finite arrays of $N\times N$ ferromagnetic nano dots 
subject to the dipole-dipole interaction are investigated for $N=2\dots 13$. 
Spin arrangements up to $N=6$ are presented, which indicate the onset of
bulk-like behavior associated with odd ($N=5$) and even ($N=6$) systems.
The effect of field misalignment on the hysteresis loops is also studied
for $N=3\cdots 6$. The area $A_N$ of the hysteresis loop is studied as a
function of $N$. We find that $A_N-A_\infty$ approximately scales as
$N^{-\frac{3}{2}}$ for $N$ odd and as $N^{-2}$ for $N$ even.
\end{abstract}
\pacs{75.40.Mg; 75.60.Ej; 75.60.Jk; 75.70.Kw}
\maketitle
\section{Introduction}
A ferromagnetic particle goes into a monodomain state if its size $D$ is below
a critical value $D_c=10\sim 100$ nm. This is due to the competition 
between the exchange and dipolar energies. Therefore, a nanoparticle in a 
monodomain state may be viewed as a giant magnetic dipole with magnetic 
moment of thousands of Bohr magnetons. For an $N\times N$ array of
well-separated nanoparticles the exchange energy is usually negligible 
in comparison with the dipolar and anisotropy and Zeeman energies.
The study of such systems is of increasing importance because of their
technological applications in data storage devices and magnetic field sensors.
As the technology of these devices moves towards higher densities of 
stored information, it requires smaller particles of magnetic media \cite{SW}, 
\cite{SSF}, for which finite size effects become relevant. In finite arrays of 
such large dipole moment particles, the dipolar field of the array becomes 
comparable with the bulk anisotropy field. Dipolar effects in such systems 
affect the static and dynamics properties of the array; and thus must be 
taken into account.

Recently, Camley and Stamps in \cite{CS1}, \cite{CS2}, \cite{CS3}
investigated the dynamics and magnetization processes of a
finite planar array of $N\times N$ ferromagnetic nano dots, for
$N=3,4,5,6$. The nano-dots were taken to interact only via the
dipole-dipole interaction, and they were subject to an
external field applied either along one side of the array or along its
diagonal. They found rather complicated hysteresis loops 
with the magnetization reversal controlled by the 
shape anisotropy induced by the array itself. We have
considered the same model, and have extended their results, for
$N=2\cdots13$. Our results for $N=3$ qualitatively agree with those of
\cite{CS1}. We find that the behavior of these systems is surprisingly
complex, both for small and for larger values of $N$, and we have studied
the approach of these systems to $N\rightarrow\infty$ behavior. 

In the present work each dot is taken to have a
radius $R_d$, thickness $d$ and a
single degree of freedom corresponding 
to the orientation of a magnet of
saturation magnetization $M_0$. We consider 
only the case of zero temperature. The dots
are arranged on a square lattice with lattice
spacing $a >2R_d$, and the dots interact only via the dipole-dipole
interaction. The equation of motion for the magnetic moment of each dot is
governed by the Landau-Lifshitz-Gilbert equation (LLG), \cite{LL} which
reads
\begin{eqnarray}
\frac{d {\bf{M}}}{dt}=\gamma
{\bf{M}}\times
{\bf{H}}_{eff} -\alpha \frac{{\bf{M}} \times
({\bf{M}}\times
{\bf{H}}_{eff})}{M_s}
\label{llg}
\end{eqnarray}
\noindent
where $\gamma$ is the gyromagnetic ratio, $\alpha$ is the damping 
coefficient, ${\bf{M}}$ is the magnetic moment of the dot and 
$M_s=|{\bf{M}}|$ is the saturation magnetization, ${\bf{H}}_{eff}$ is the 
average effective magnetic field acting on the dot. 
The average effective magnetic field acting on the
$i$-th dot is due to the applied external field, the dipolar
fields, and the anisotropy field
\begin{eqnarray}
{\bf{H}}_{eff}^i = H_0 \cos \theta \hat{x} + H_0 \sin \theta \hat{y} 
-{\bf{H}}_{dip}^i + 2 K_1 \frac{m_z^i}{M^2} \hat{z}.
\label{heff}
\end{eqnarray} 
Here the dipole field acting on the $i$-th dot due to all other dots in 
the array is given by 
\begin{eqnarray}
{\bf{H}}_{dip}^i =\sum_{j\ne i}\frac{{\bf{M}}_j}{r_{ij}^3} -3
\frac{({\bf{M}}_j\cdot{\bf{r}}_{ij}){\bf{r}}_{ij}}{r_{ij}^5}
\end{eqnarray}

The choice of anisotropy field is determined by the shape of the dot,
which in our problem is directed along the symmetry axis of the
(cylindrical) dots. We divide both sides of Eq.(\ref{llg}) by 
($\gamma M_s^2$) and define a dimensionless time variable 
$\tau=\gamma M_s t$. The LLG in these reduced units becomes

\begin{eqnarray}
\frac{d{\bf{m}}}{d\tau}={\bf{m}}\times {\bf{h}}_{eff}
-\frac{\alpha}{\gamma} {\bf{m}} \times ({\bf{m}}\times {\bf{h}}_{eff})
\label{llg2}
\end{eqnarray}
where ${\bf{m}}=\frac{{\bf{M}}}{M_s}$ and ${\bf{h}}_{eff}=
\frac{{\bf{H}}_{eff}}{M_s}$. We employ a system of
units where magnetic fields are given in units of 
$M_s$ and distances are given in units of the array's lattice spacing $a$. 
The strength of the dipole field is characterized by $h_{dip}=
\frac{\pi R_d^2 d}{a^3}$, which is the ratio of the dot volume to the 
volume of a cubic cell with side $a$. For all arrays studied in 
this work, we take $h_{dip}=0.5$. 

This article is organized as follows. Section II presents the 
numerical techniques and method of time integration. Section III  
presents an extensive discussion of magnetization processes and 
hysteresis for arrays of $N\times N$ nano dots ($N=2, 3, \cdots, 13$) 
when the external field is applied along one side of the array. 
Section IV considers the effects on the hysteresis loop of   
misalignment of the external field. Section V considers the
relationship between the area of the hysteresis loop and $N$. 
A brief summary is given in section VI.   

\section{Numerical Techniques}
We have employed two different approaches to study the magnetization
processes of our $N\times N$ arrays of nano-dots. The first approach uses
the second rank Runge-Kutta (RK) algorithm with fixed time step to
integrate the LLG equation in reduced units. The second approach uses the
``greedy algorithm`` to find the stable final state of the array.
In both approaches the calculations were done using FORTRAN and Mathematica
languages \cite{walf}. The two approaches yielded similar results.
 
In the RK approach, the integration is done using a fixed time step $\Delta
\tau =5\times 10^{-3}$ and a damping coefficient $\frac{\alpha}{\gamma}=0.6$, 
with an initial state in which the magnetic moment of the dot is randomly 
generated. The time integration proceeds until a stable final state is 
reached. The stability of the final state is checked by changes in the 
total energy of the system. Iterations are stopped when the difference 
between the total energy of the system from the ($n+1$)-th iteration and 
that of the $n$-th iteration is of the order of $\Delta E_n=10^{-5}$ in 
units of $M_s^2 a^3$. Our solutions converged after almost $10^3$ iterations.

The greedy algorithm approach assumes the dots to be aligned
along the direction of their total local field. In the initial state each
dot is chosen to point randomly. Next, all components of the total local
field are calculated for each dot. Finally each component of the field is
normalized by the magnitude of the total local field to yield the new dot
magnetization orientation. The final state of the $n$-th iteration is used
as an initial one for the next iteration until the solution is converged.
The convergence of the final state is checked in manner similar to that
used in the first approach. Note that the greedy algorithm can be obtained
from the LLG equation in the limit where the damping coefficient is very
large. We find that the LLG approach converges faster than the greedy
algorithm. This is probably related to a phenomenon known in RLC
circuits, where a critically-damped circuit approaches equilibrium faster
than a circuit with a larger amount of damping.

Calculation of the dipolar field at the $i$-th dot is the most
computationally time-consuming aspect of both approaches since it
requires summation of the dipole fields from all other dots in the array.
However, due to the relatively small sizes of our dot arrays  
this calculation is performed rather quickly.

\section{Hysteresis Loop and External Field Orientation Effect}
\begin{figure}[ht]
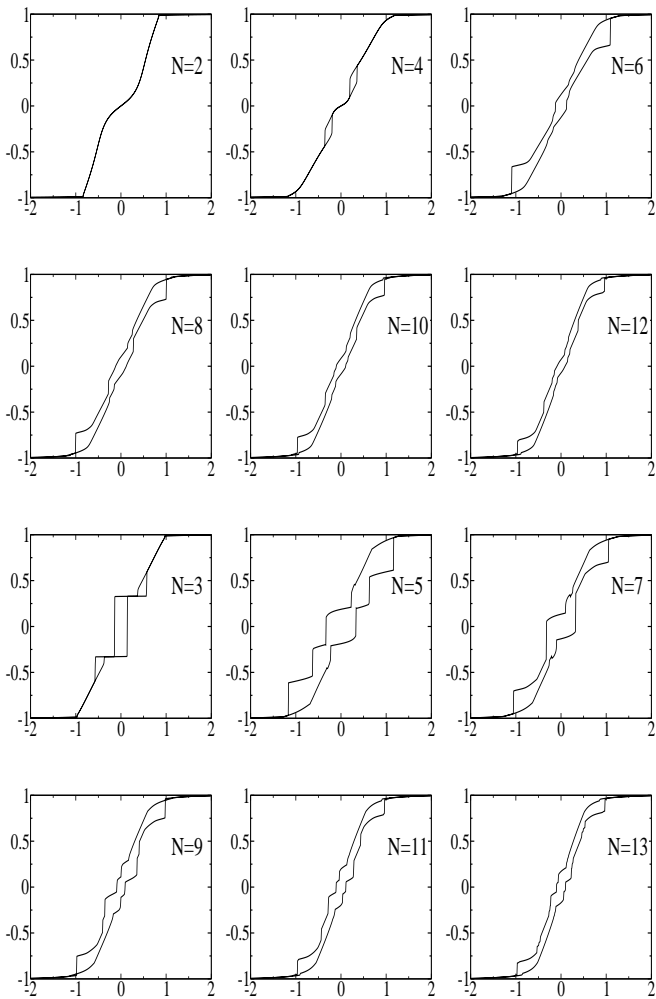

  \centering
  \includegraphics[angle=0,width=1.1in,totalheight=1.1in]{hyst1x2x2.eps}
  \hfill
  \includegraphics[angle=0,width=1.1in,totalheight=1.1in]{hyst1x4x4.eps}
  \hfill
  \includegraphics[angle=0,width=1.1in,totalheight=1.1in]{hyst1x6x6.eps}\\
  \vspace{0.25in}
  \includegraphics[angle=0,width=1.1in,totalheight=1.1in]{hyst1x8x8.eps}
  \hfill
  \includegraphics[angle=0,width=1.1in,totalheight=1.1in]{hyst1x10x10.eps}
  \hfill
  \includegraphics[angle=0,width=1.1in,totalheight=1.1in]{hyst1x12x12.eps}\\
  \vspace{0.25in}
  \includegraphics[angle=0,width=1.1in,totalheight=1.1in]{hyst1x3x3.eps}
  \hfill
  \includegraphics[angle=0,width=1.1in,totalheight=1.1in]{hyst1x5x5.eps}
  \hfill
  \includegraphics[angle=0,width=1.1in,totalheight=1.1in]{hyst1x7x7.eps}\\
  \vspace{0.25in}
  \includegraphics[angle=0,width=1.1in,totalheight=1.1in]{hyst1x9x9.eps}
  \hfill
  \includegraphics[angle=0,width=1.1in,totalheight=1.1in]{hyst1x11x11.eps}
  \hfill
  \includegraphics[angle=0,width=1.1in,totalheight=1.1in]{hyst1x13x13.eps}
  \caption{Hysteresis loops $M(H)$, both in units of $M_s$, for weakly
    coupled arrays of $N\times N$ ferromagnetic nano dots.  The external
    field is applied along the $y$-axis, which coincides with one side of
    the array. The top two rows are for $N$ even and the lower two are for
    $N$ odd.}
  \label{fig1}	
\end{figure}

Hysteresis loops $M(H)$ for $N\times N$ arrays of nano dots subject to 
an external magnetic field applied along one side have been calculated.
Initially, a strong external field $H_{0}$ is applied to the array until 
saturation. Then the field is decreased to $-H_{0}$, and then is increased 
again up to $H_{0}$. We have taken $H_{0} =2M_s$, and a fixed field-step of 
$\Delta H =2\times 10^{-3}M_s$ is used to simulate the sweeping process. 
For each value of the external field the system was iterated until a 
stable final state is reached. As seen in Fig.~\ref{fig1}, where $M(H)$ is 
plotted (both in units of $M_s$), the odd and even $N$ arrays have 
somewhat different behaviors, especially for small $N$. One aspect of this is 
that the odd-$N$ systems display magnetization jumps as the field changes. 
The odd-$N$ and even-$N$ behavior becomes similar for larger values of $N$, 
something we study in a later section. 

In Fig.~\ref{fig1}, the angle of the field to one of the sides (the $y$-axis) 
is taken to be $\theta=0$. Experimentally, however, field misalignment is 
almost inevitable, so that we have also studied this phenomenon.

\begin{figure}[h]
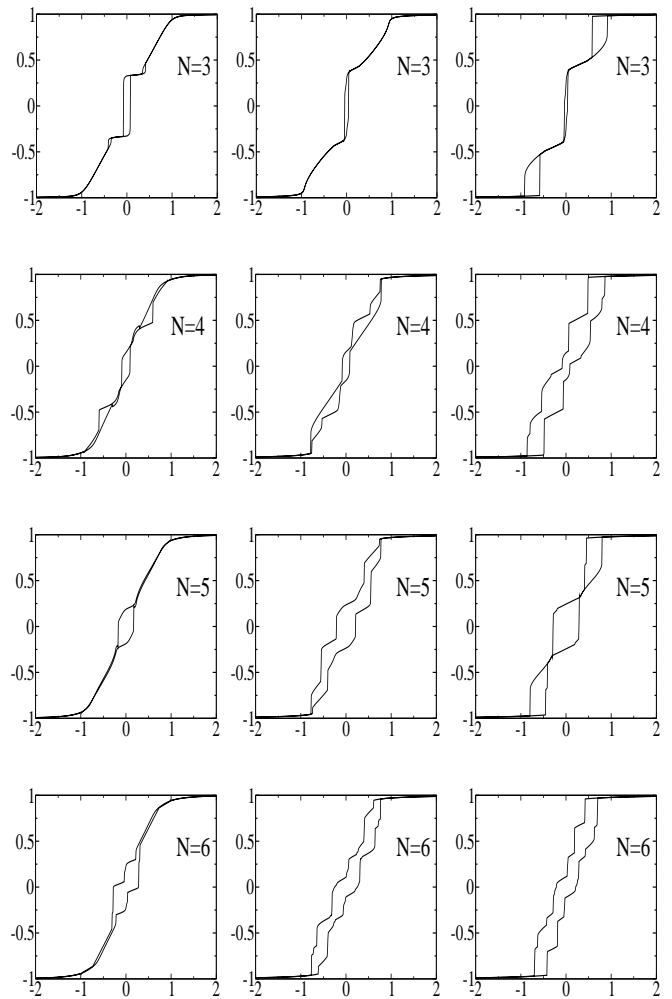

  \centering
  \hfill
  \includegraphics[angle=0,width=1.1in,totalheight=1.1in]{3x3angle05.eps}
  \hfill
  \includegraphics[angle=0,width=1.1in,totalheight=1.1in]{3x3angle30.eps}
  \hfill
  \includegraphics[angle=0,width=1.1in,totalheight=1.1in]{3x3angle45.eps}\\
  \vspace{0.25in}
  \includegraphics[angle=0,width=1.1in,totalheight=1.1in]{4x4angle05.eps}
  \hfill
  \includegraphics[angle=0,width=1.1in,totalheight=1.1in]{4x4angle30.eps}
  \hfill
  \includegraphics[angle=0,width=1.1in,totalheight=1.1in]{4x4angle45.eps}\\
  \vspace{0.25in}
  \includegraphics[angle=0,width=1.1in,totalheight=1.1in]{5x5angle05.eps}
  \hfill
  \includegraphics[angle=0,width=1.1in,totalheight=1.1in]{5x5angle30.eps}
  \hfill
  \includegraphics[angle=0,width=1.1in,totalheight=1.1in]{5x5angle45.eps}\\
  \vspace{0.25in}
  \includegraphics[angle=0,width=1.1in,totalheight=1.1in]{6x6angle05.eps}
  \hfill
  \includegraphics[angle=0,width=1.1in,totalheight=1.1in]{6x6angle30.eps}
  \hfill
  \includegraphics[angle=0,width=1.1in,totalheight=1.1in]{6x6angle45.eps}
  \caption{Hysteresis loops $M(H)$, both in units of $M_s$, for $N=3\cdots 6$
    arrays for the field at $\theta=5^\circ, 30^\circ$ and $45^\circ$ going
    from left to right.}
  \label{fig88}
\end{figure}

Fig.~\ref{fig88} shows results for $N=3\dots 6$ and angles
$\theta=5^\circ$, $30^\circ$, and $45^\circ$. (We present only some of
the more representative results; angles between 0 and $45^\circ$ were
studied in $5^\circ$ increments.) 

Comparison of
Fig.~\ref{fig1} with Fig.~\ref{fig88} for $\theta=5^\circ$ shows that 
a small misalignment of the applied field can change the hysteresis loop
drastically.

For $N= 3$, Fig.\ref{fig88} shows that the central part of
the hysteresis loop shrinks as $\theta$ increases. For $\theta=45^\circ$,
the central part almost disappears completely, and new small loops start
to develop away from the center of the hysteresis loop. Our results for
$N=3$ agree with those given in \cite{CS1}, which studied the cases
$\theta=0^\circ$ and $45^\circ$. 

For $N=4$, at $\theta=0^\circ$ there is no central loop, but there is 
a prominent loop at finite field. On the other hand, at $\theta=5^\circ$, 
there is a central loop, and the finite
field structure becomes rather complex. Further increase of 
$\theta$ leads to a filling out and connecting of various subloops. 
Also, note the appearance of jumps for non-zero $\theta$.  

For $N=5$, a small field misalignment has an enormous effect, 
at $\theta=5^\circ$ shrinking the loop to a relatively small central 
region. As $\theta$ increases, the central loop grows, but the loop for 
$\theta=45^\circ$ pinches off to yield three subloops, as for $N=3$.   

For $N=6$, again a small field misalignment has an enormous effect, 
at $\theta=5^\circ$ shrinking the loop to a relatively small central 
region. As $\theta$ increases, the central loop grows, but in contrast 
to $N=5$, the loop for $\theta=45^\circ$ does not pinch off, and closely 
resembles the loop for $N=4$. 

These different types of behavior indicates that these are complex
systems, for which it is difficult to make generalizations.

\section{Hysteresis and Even-Odd Signature in Finite Array of Nano-Dots}
\begin{figure}[h]
  \centering
  \includegraphics[angle=0,width=0.9in,totalheight=1.0in]{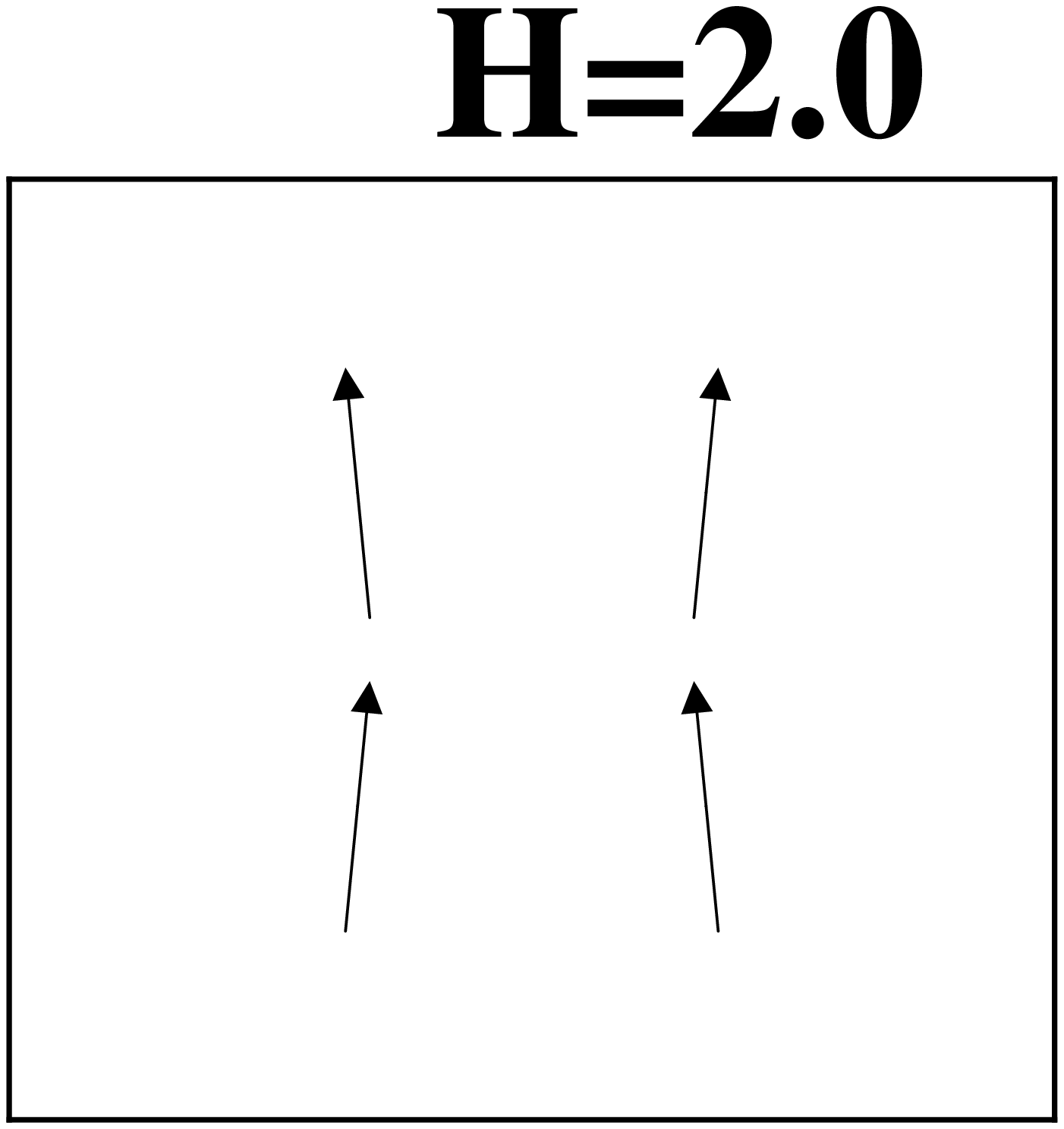}
  \hfill
  \includegraphics[angle=0,width=0.9in,totalheight=1.0in]{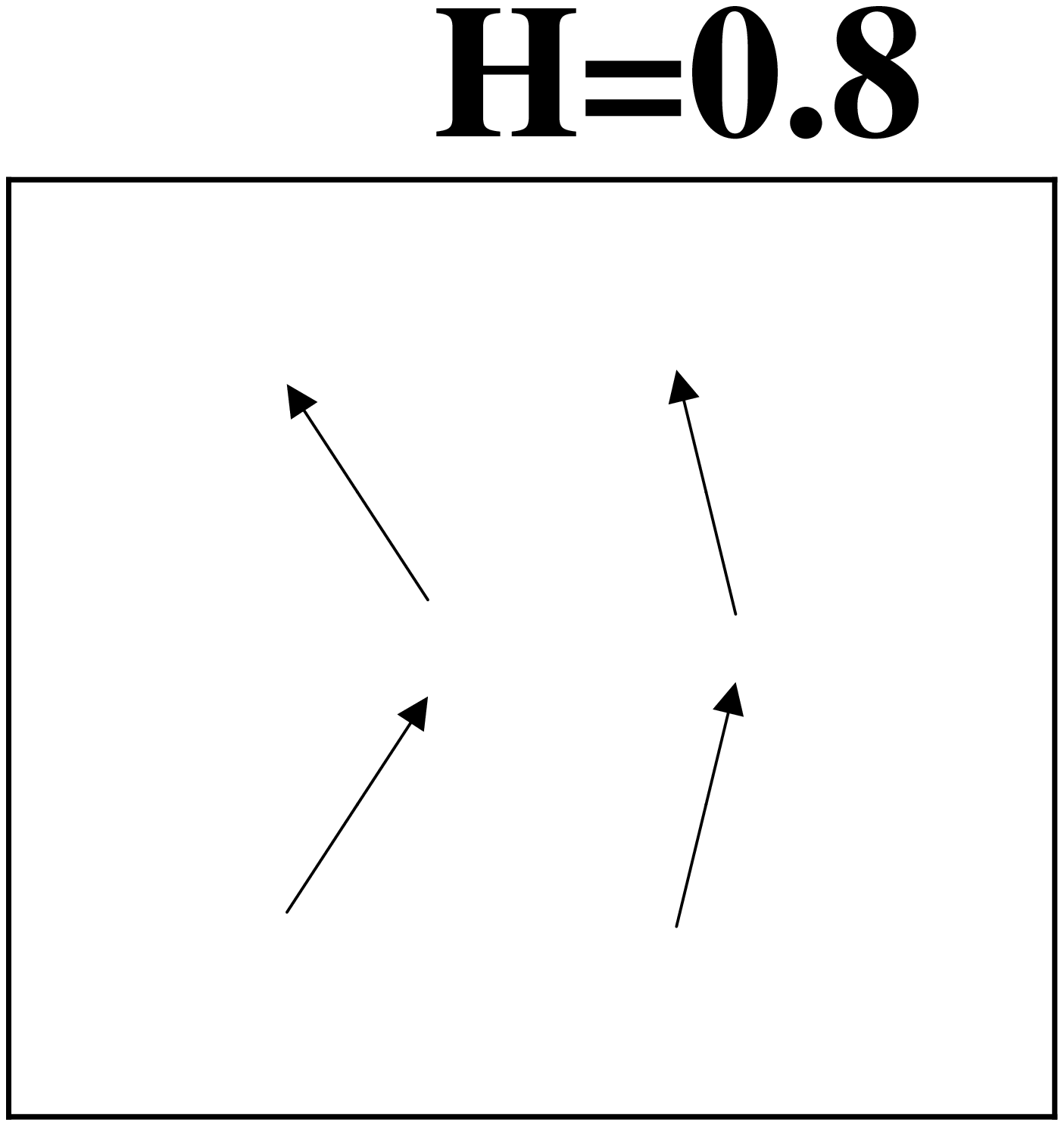}
  \hfill
  \includegraphics[angle=0,width=0.9in,totalheight=1.0in]{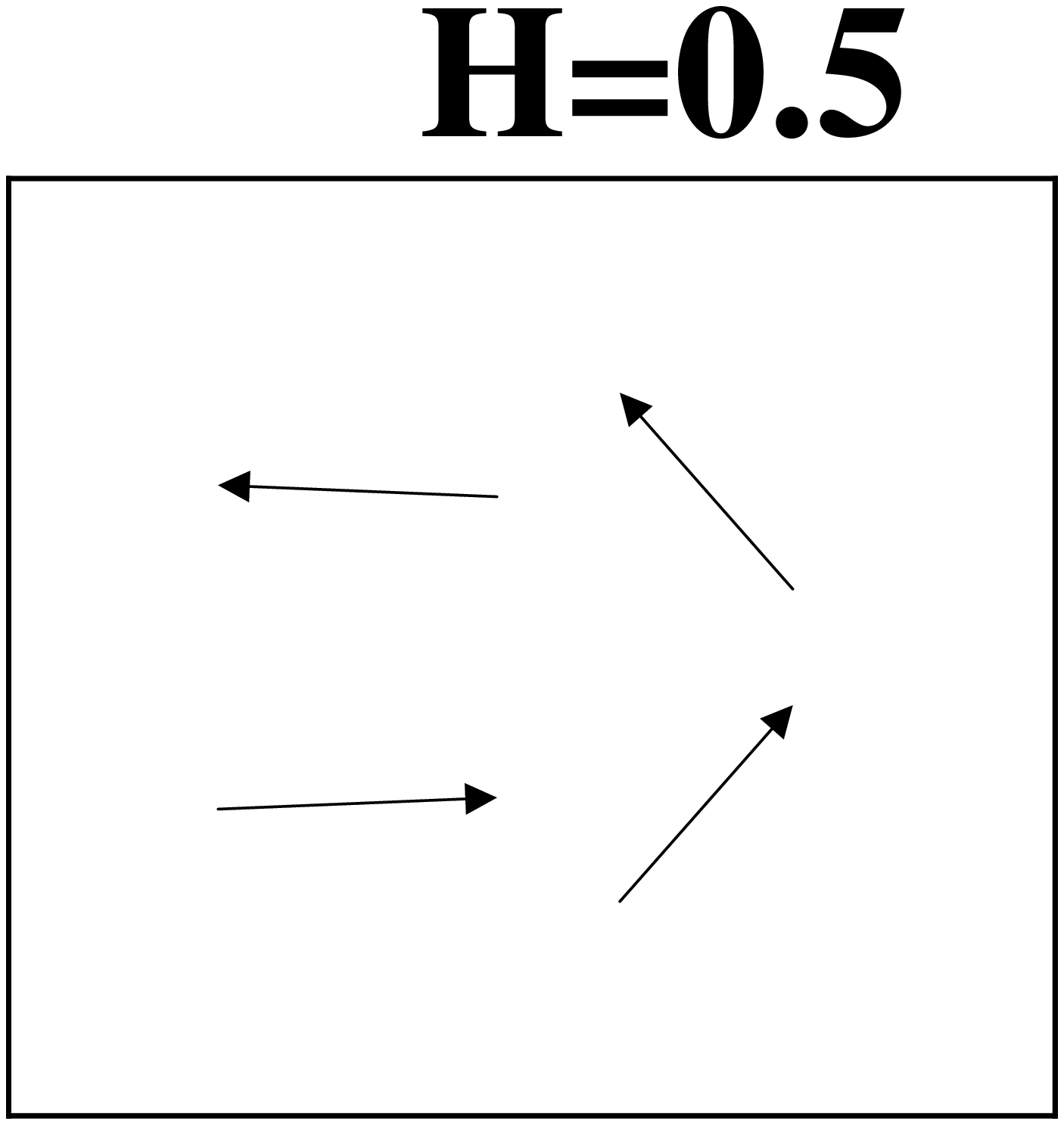}\\
  \vspace{0.15in}
  \includegraphics[angle=0,width=0.9in,totalheight=1.0in]{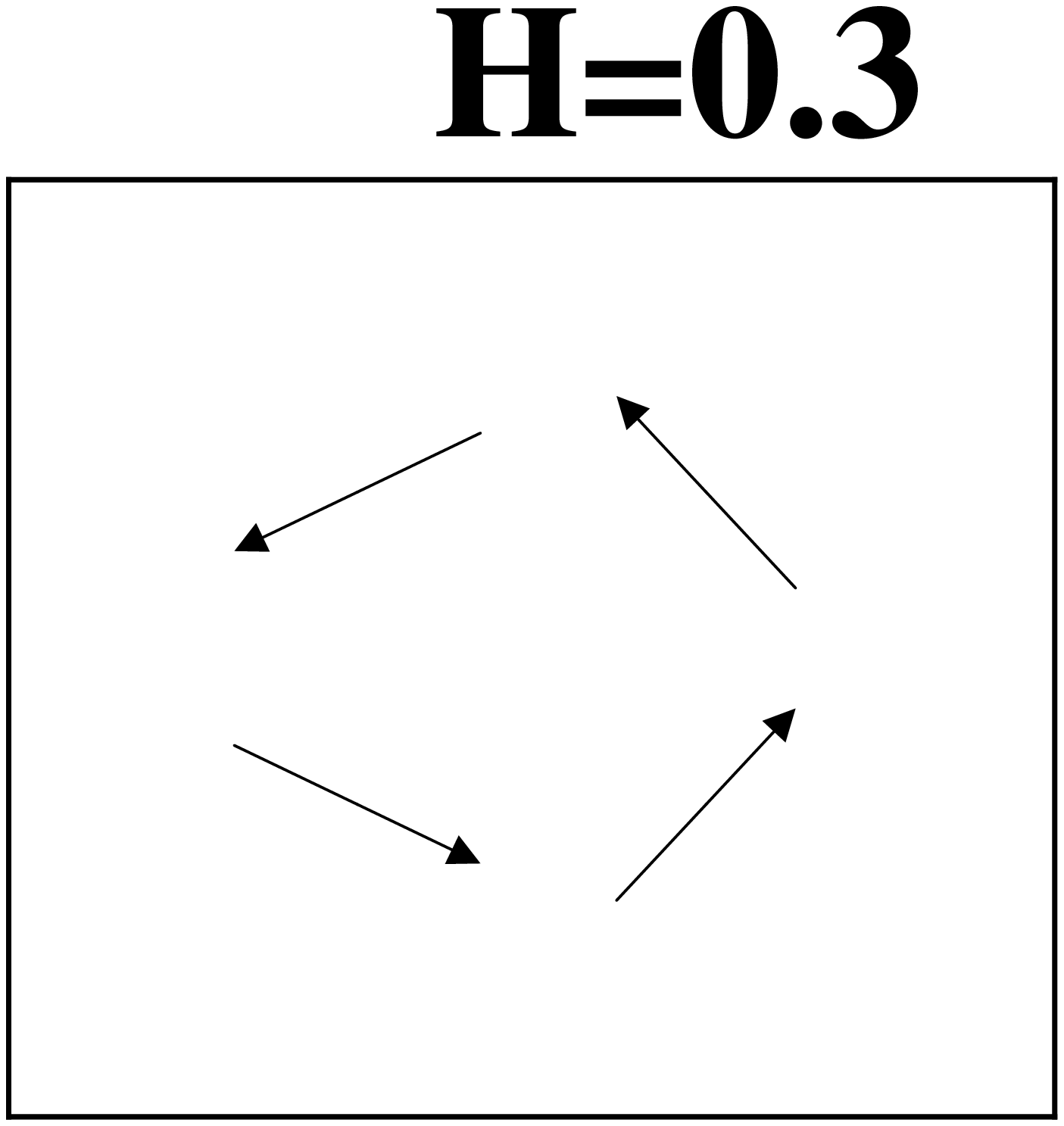}
  \hfill
  \includegraphics[angle=0,width=0.9in,totalheight=1.0in]{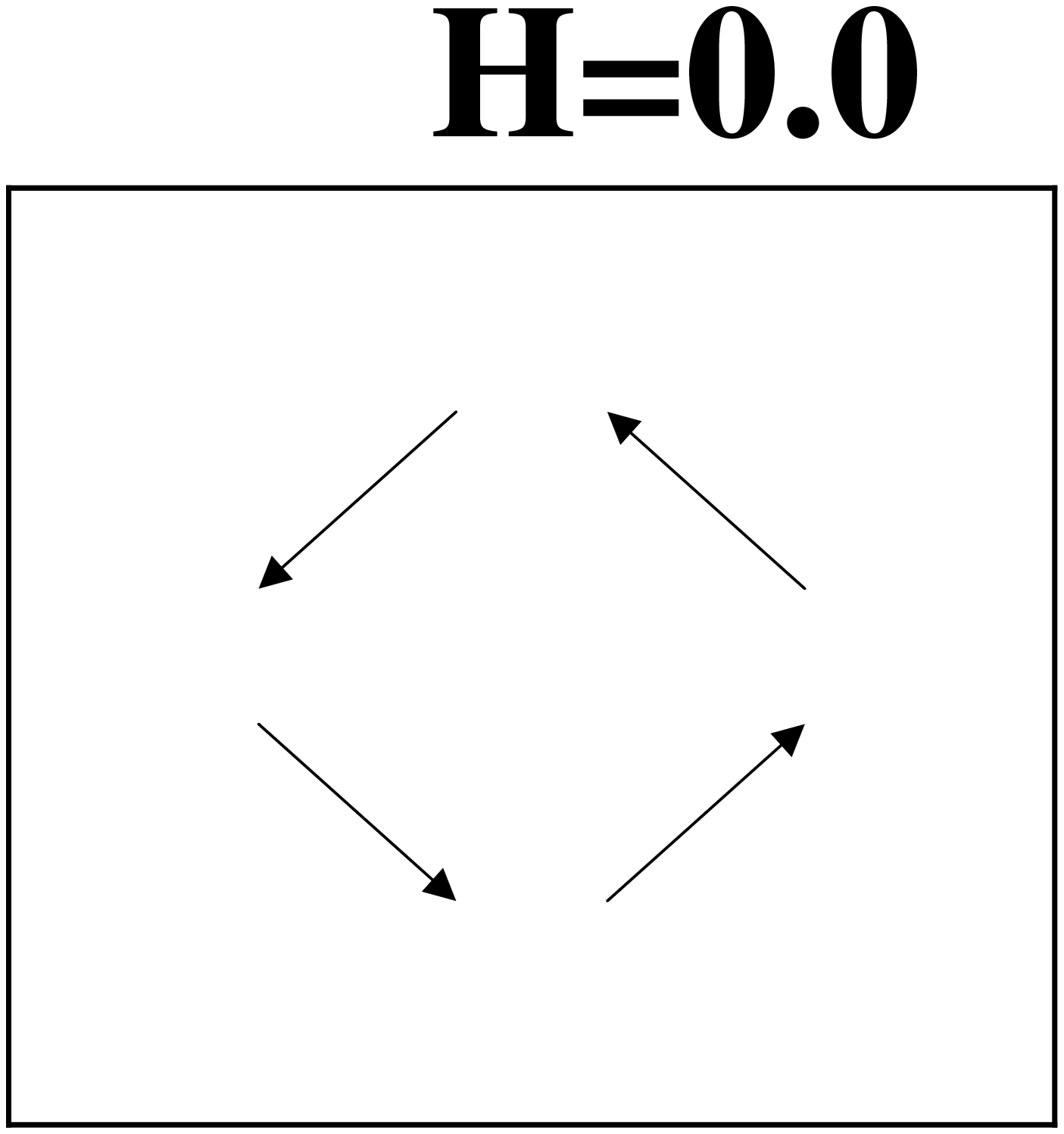}
  \hfill
  \includegraphics[angle=0,width=0.9in,totalheight=1.0in]{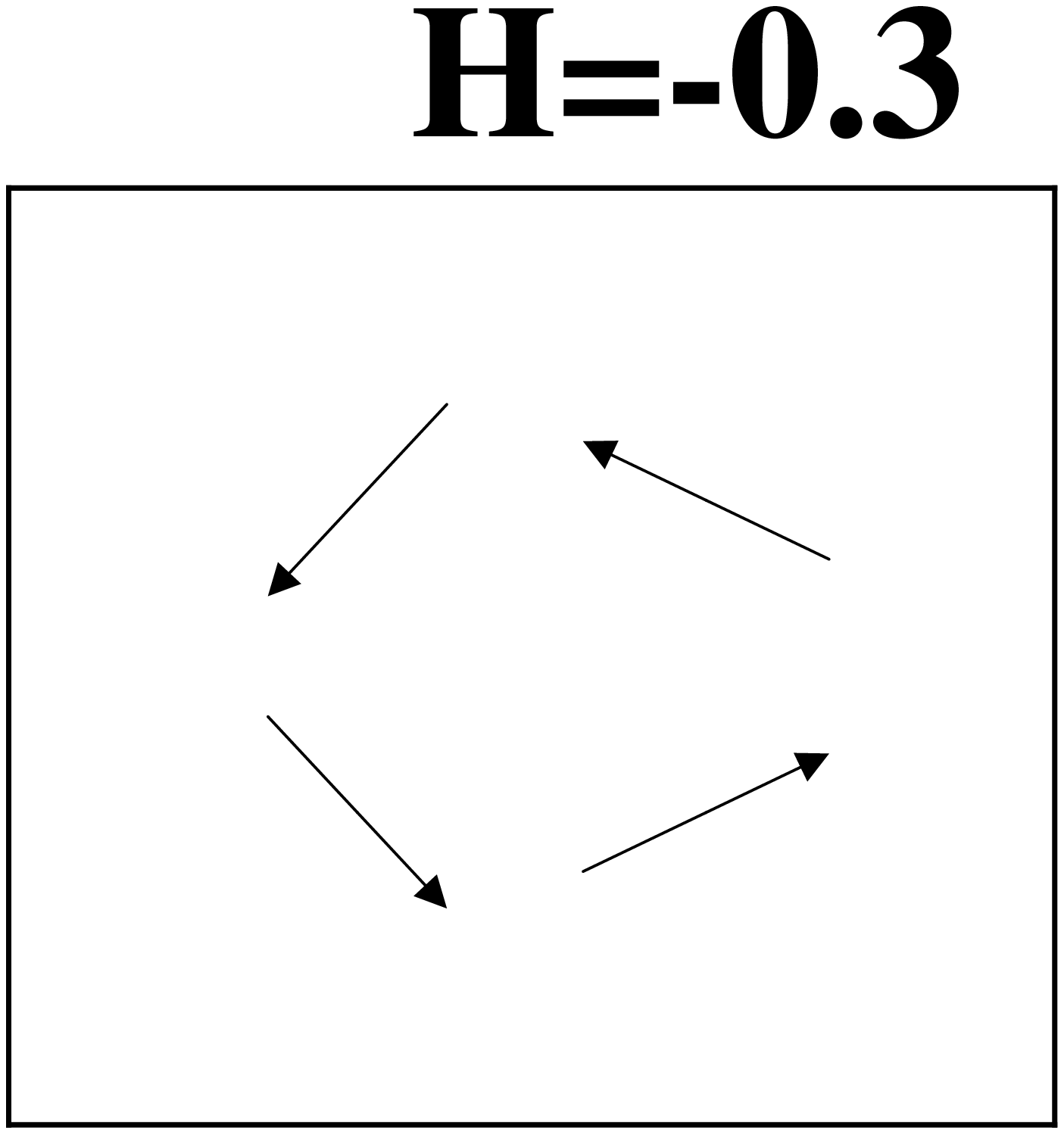}\\
  \vspace{0.15in}
  \includegraphics[angle=0,width=0.9in,totalheight=1.0in]{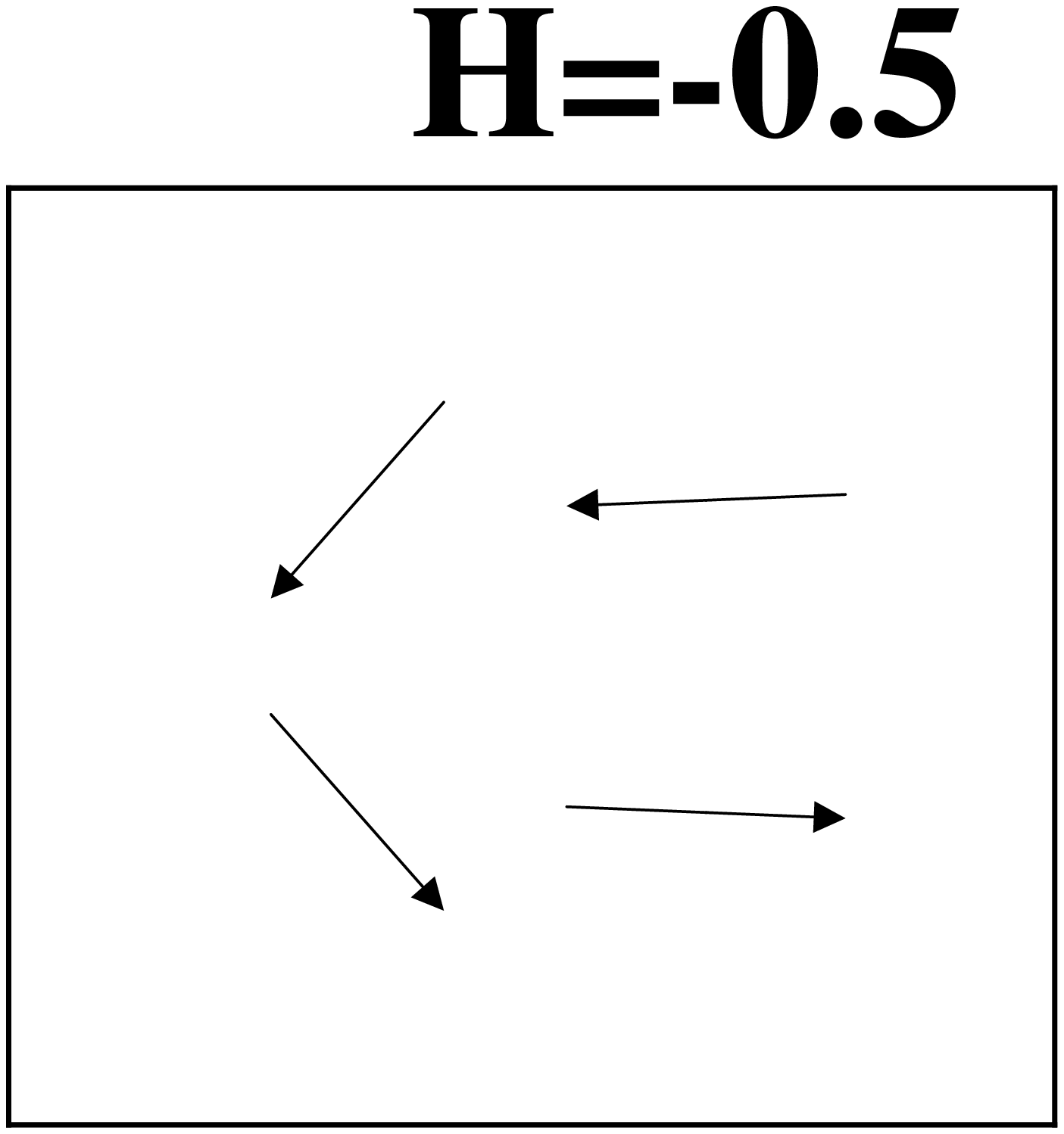}
  \hfill
  \includegraphics[angle=0,width=0.9in,totalheight=1.0in]{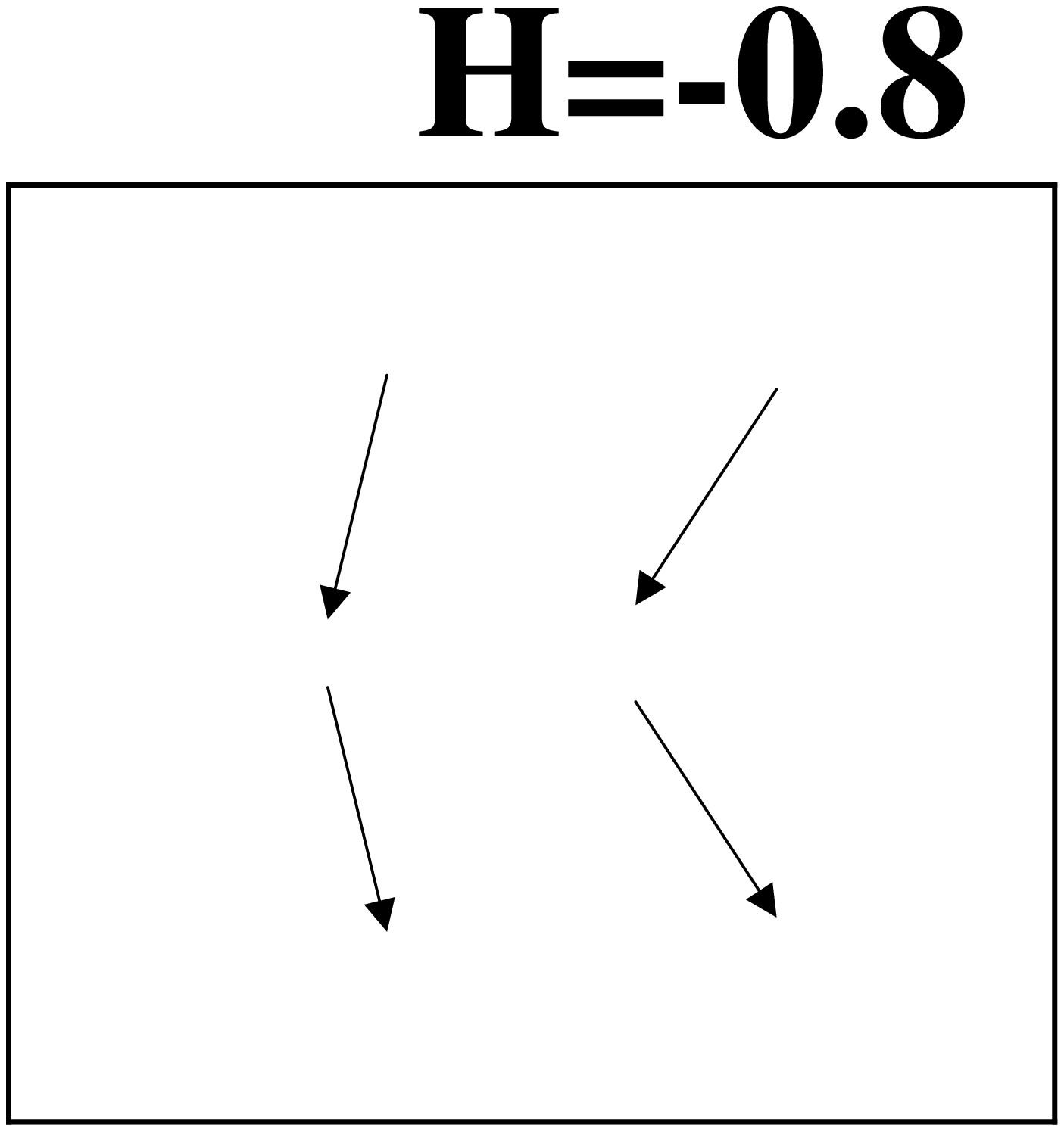}
  \hfill
  \includegraphics[angle=0,width=0.9in,totalheight=1.0in]{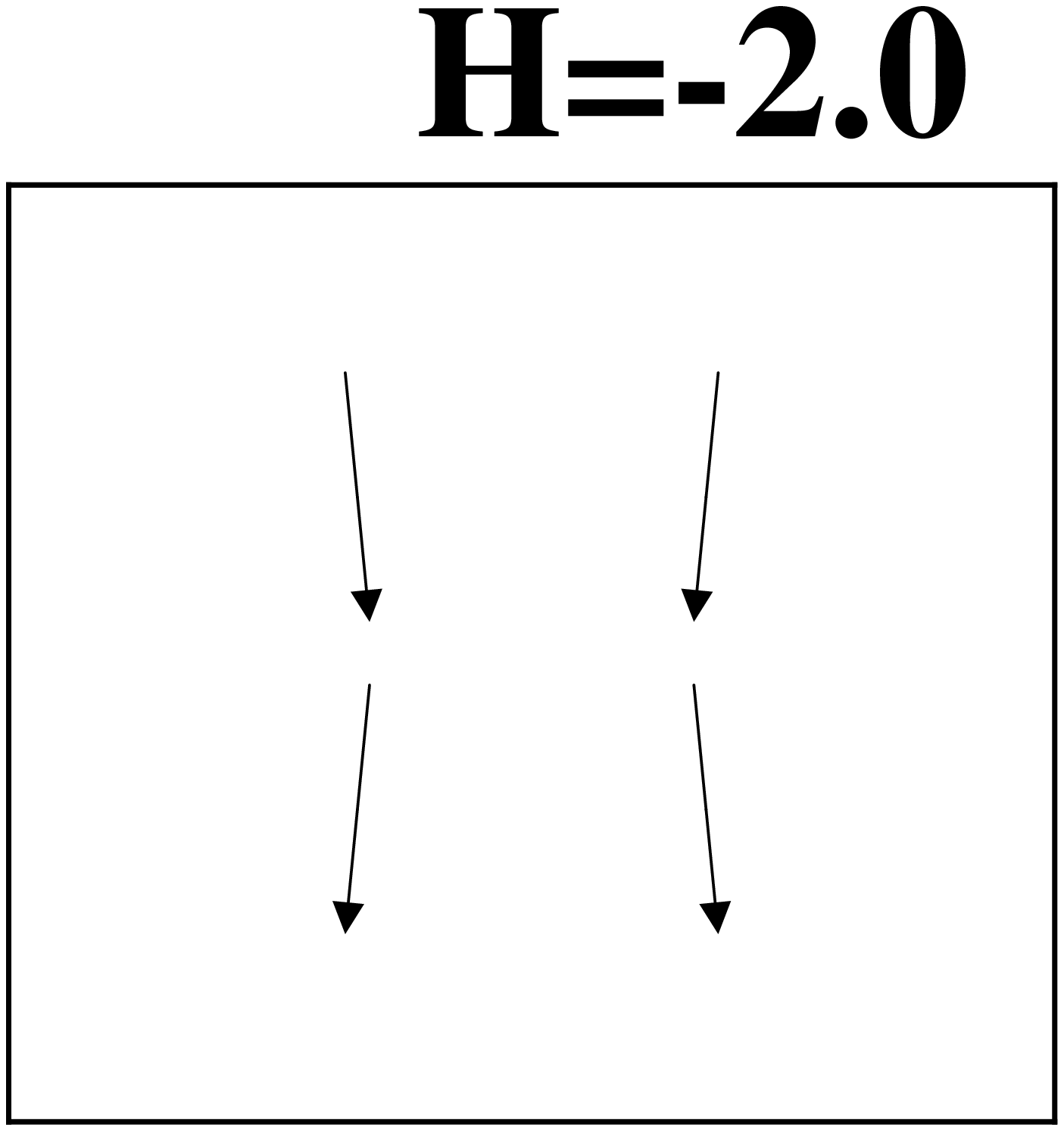}
  \caption{Spin arrangements for an array of $2\times 2$ ferromagnetic 
    nano dots in external magnetic field.}
  \label{fig2}
\end{figure}

\begin{figure}[h]
  \centering
  \includegraphics[angle=0,width=0.9in,totalheight=1.0in]{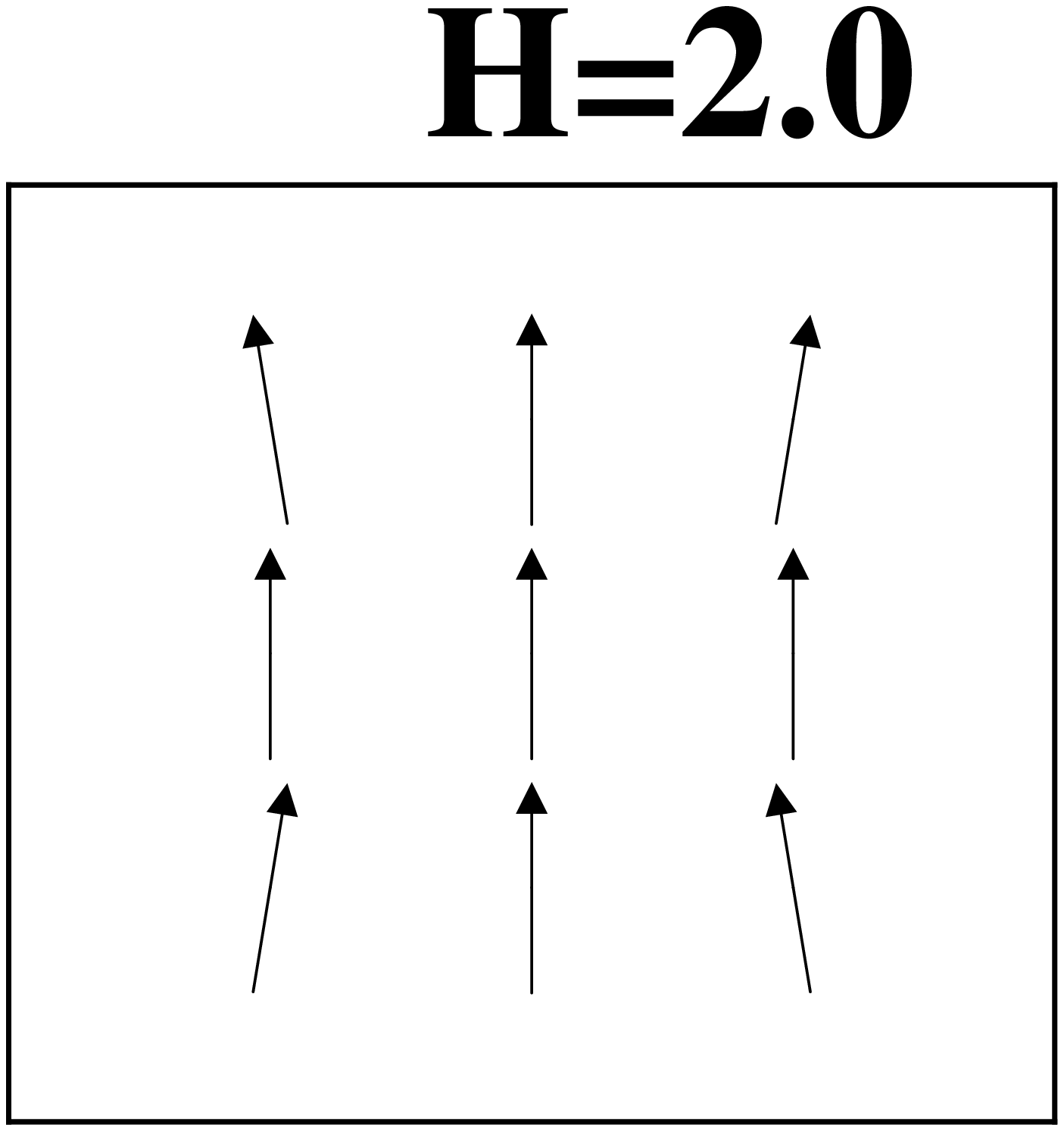}
  \hfill
  \includegraphics[angle=0,width=0.9in,totalheight=1.0in]{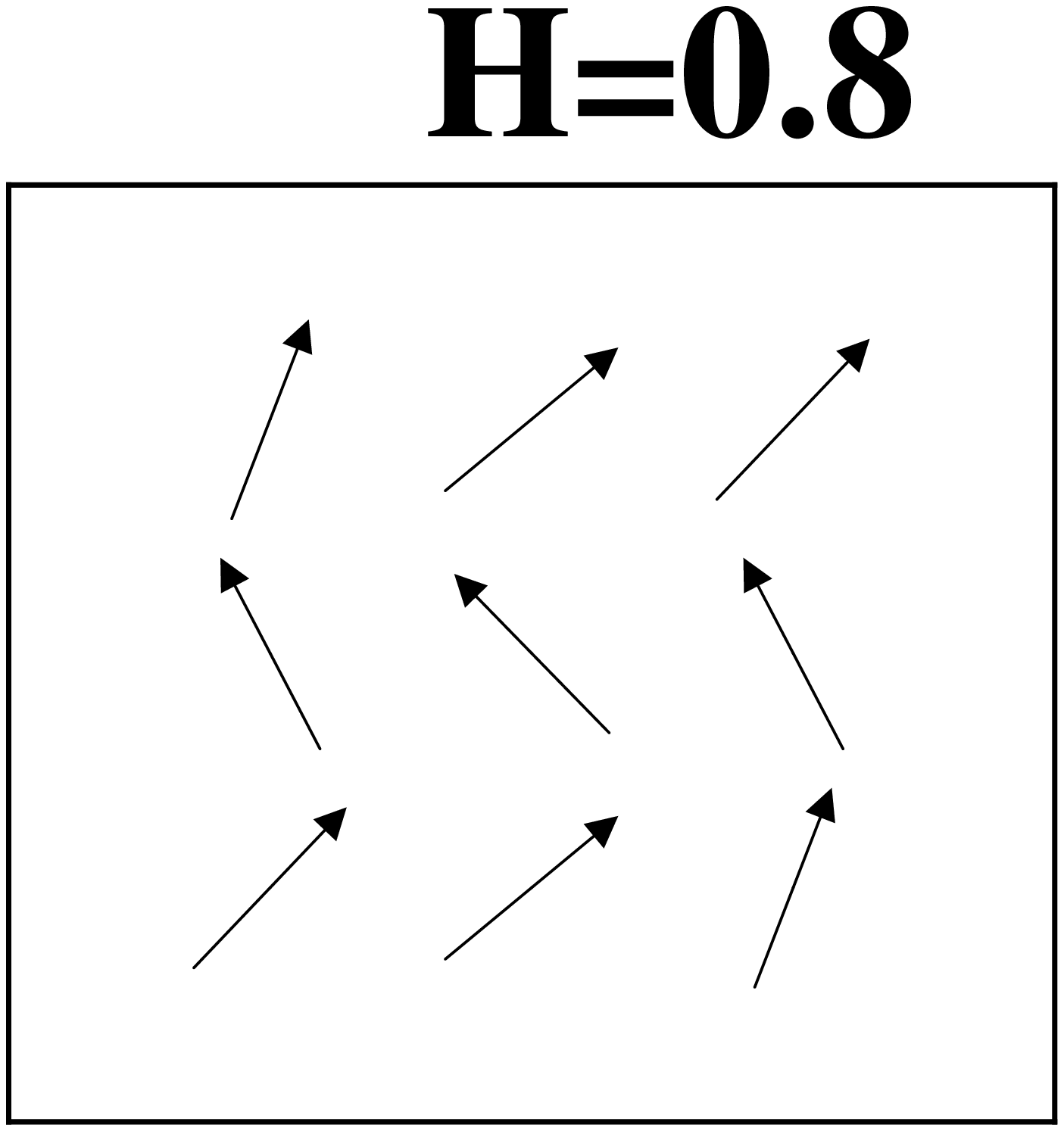}
  \hfill
  \includegraphics[angle=0,width=0.9in,totalheight=1.0in]{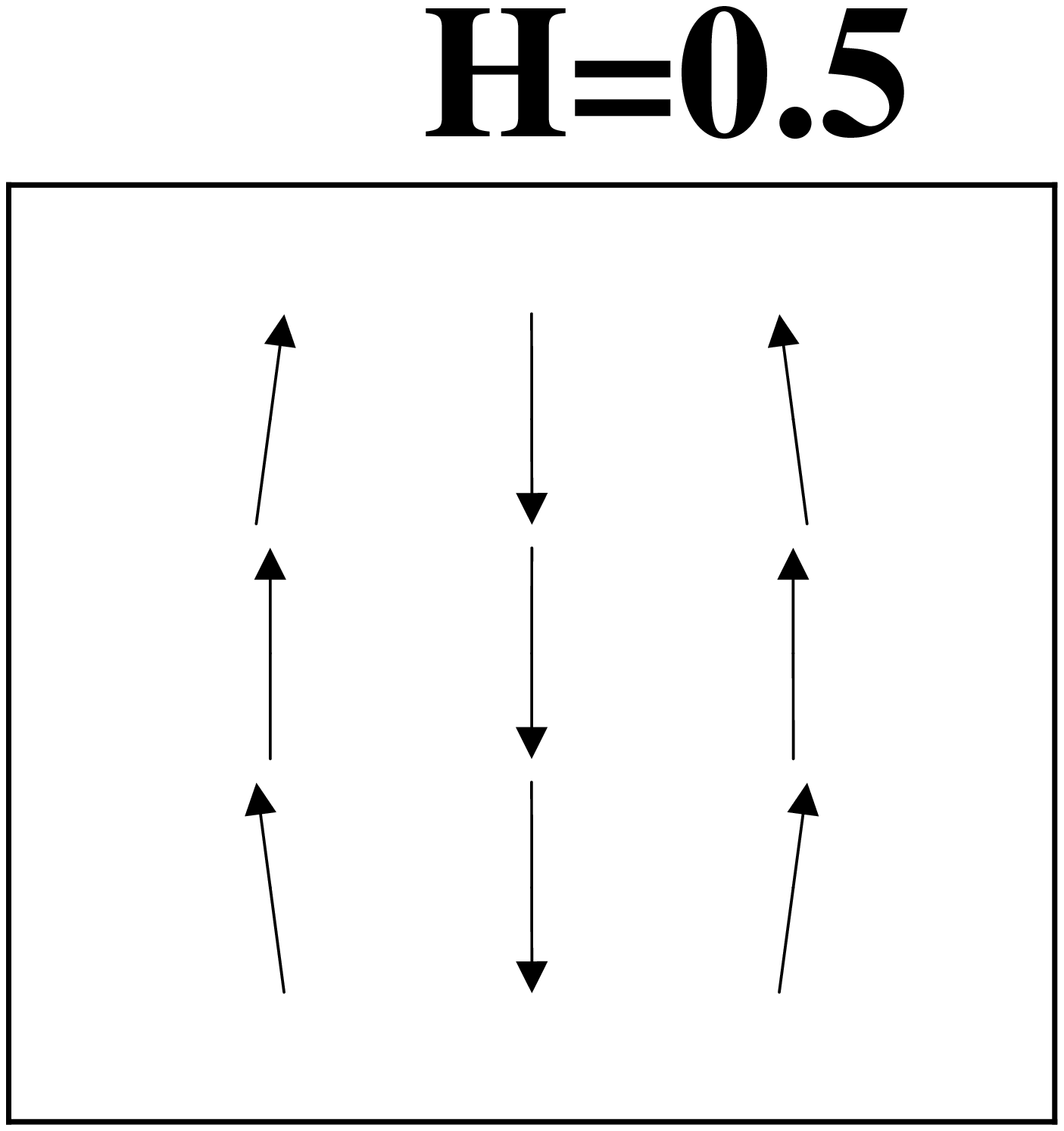} \\
  \vspace{0.15in}
  \includegraphics[angle=0,width=0.9in,totalheight=1.0in]{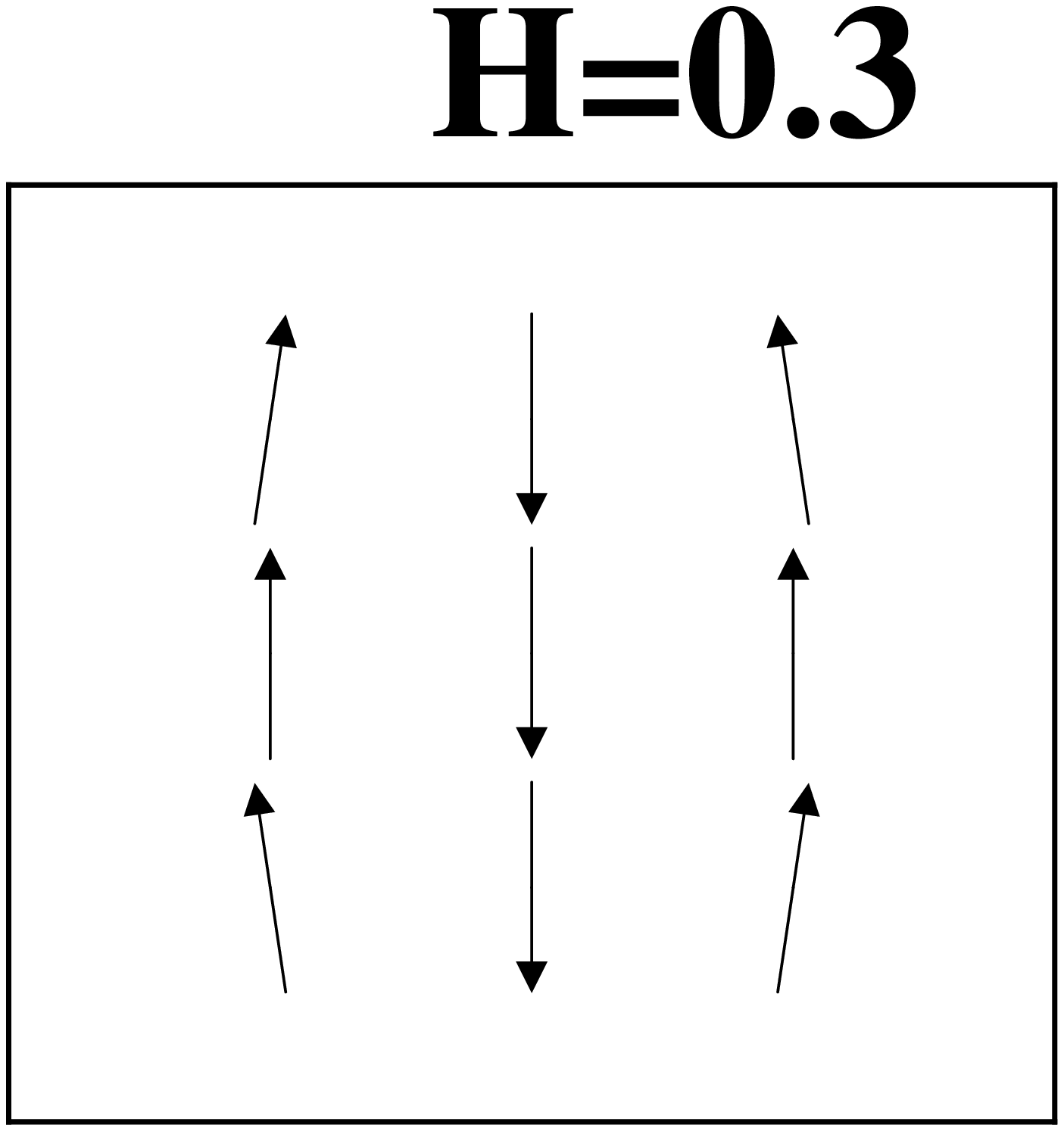}
  \hfill
  \includegraphics[angle=0,width=0.9in,totalheight=1.0in]{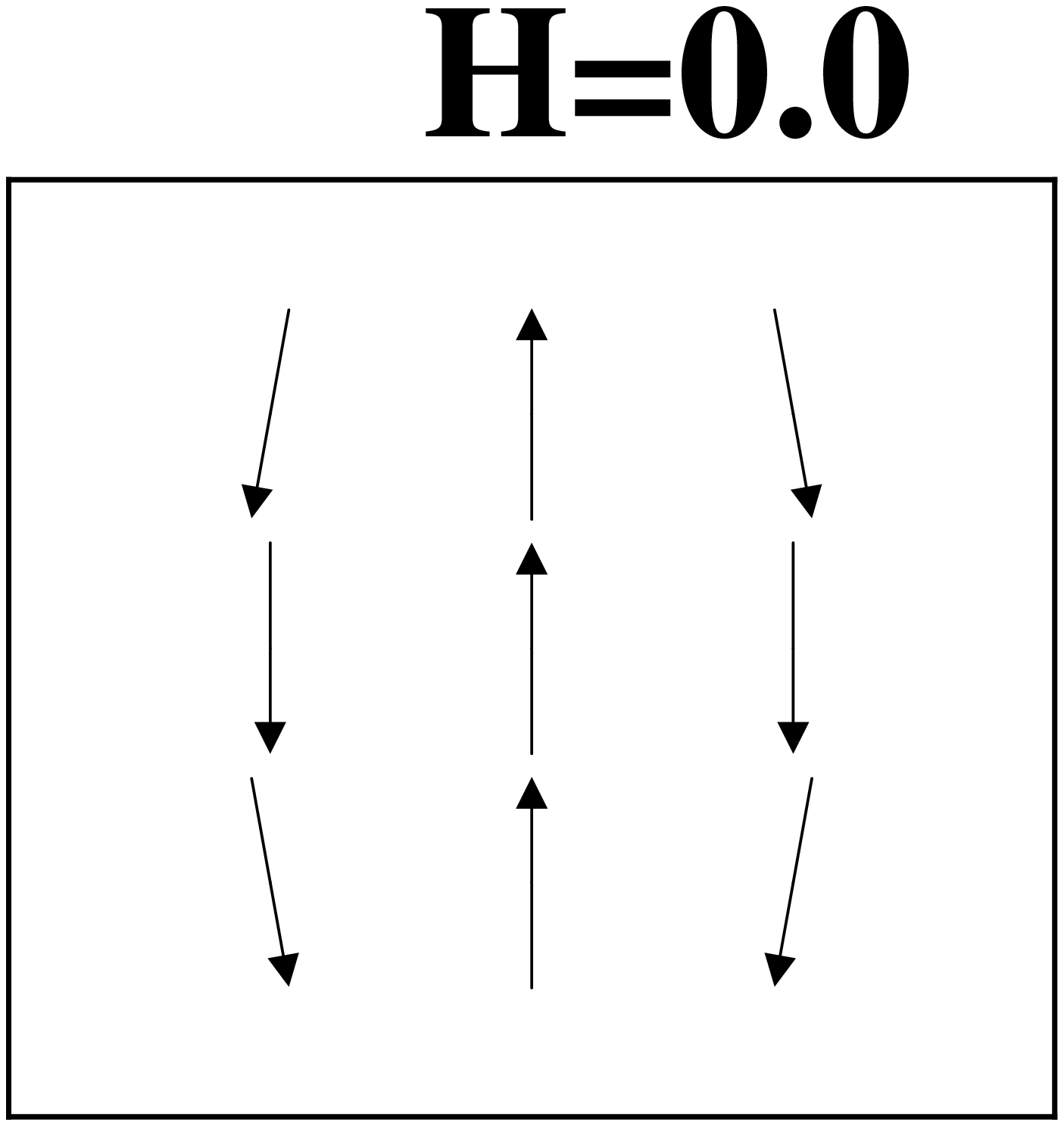}
  \hfill
  \includegraphics[angle=0,width=0.9in,totalheight=1.0in]{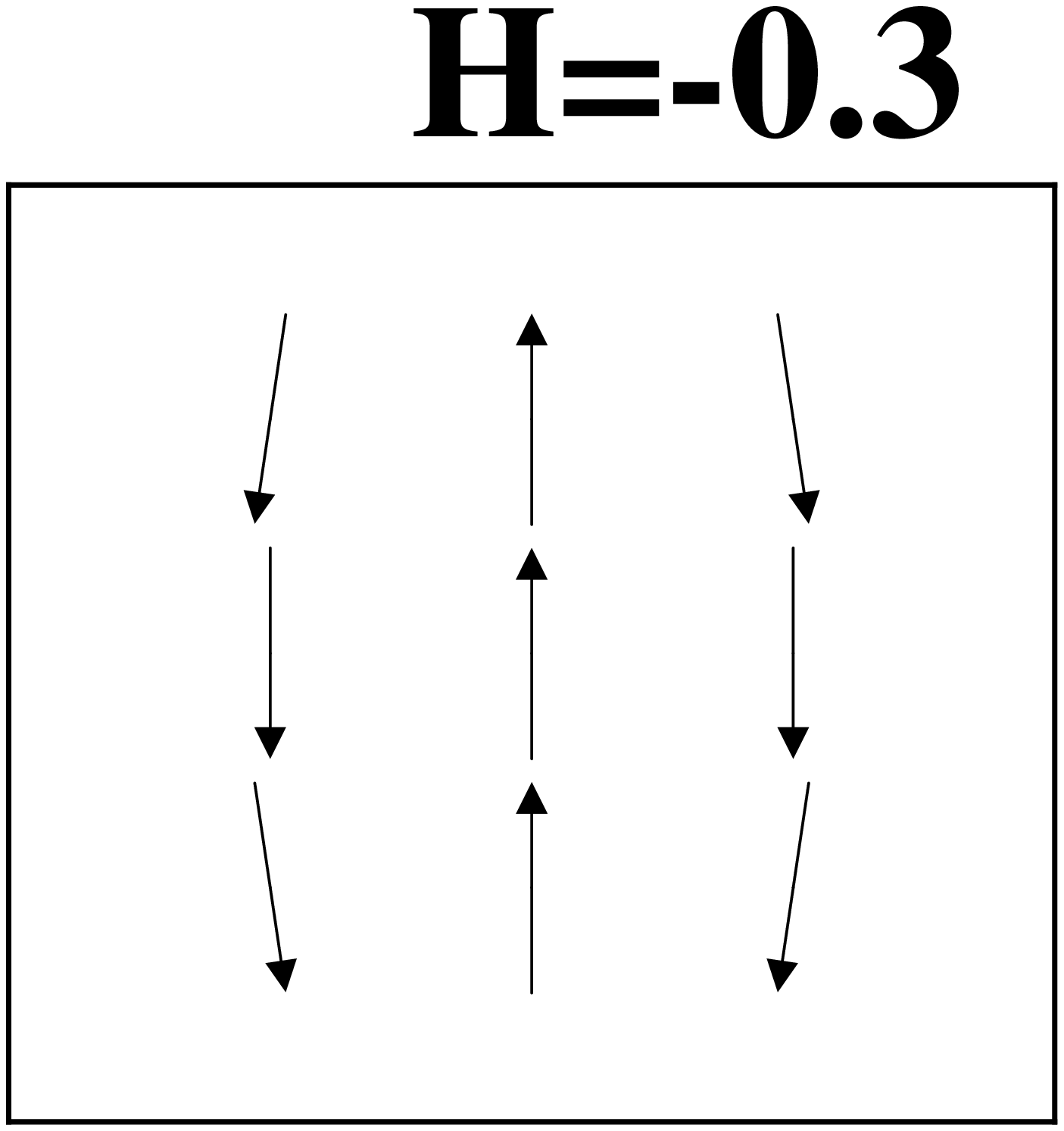}\\
  \vspace{0.15in}
  \includegraphics[angle=0,width=0.9in,totalheight=1.0in]{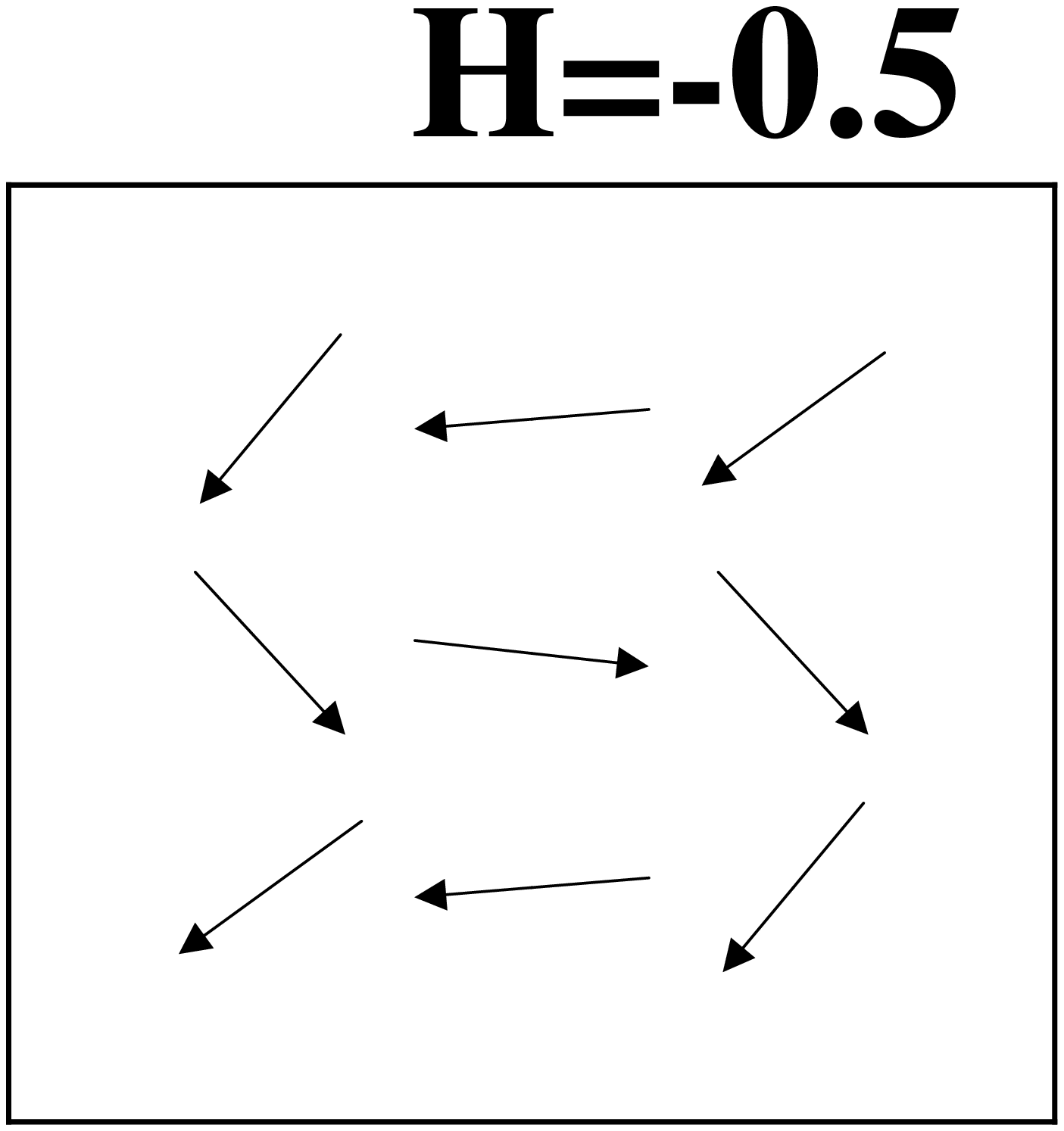}
  \hfill
  \includegraphics[angle=0,width=0.9in,totalheight=1.0in]{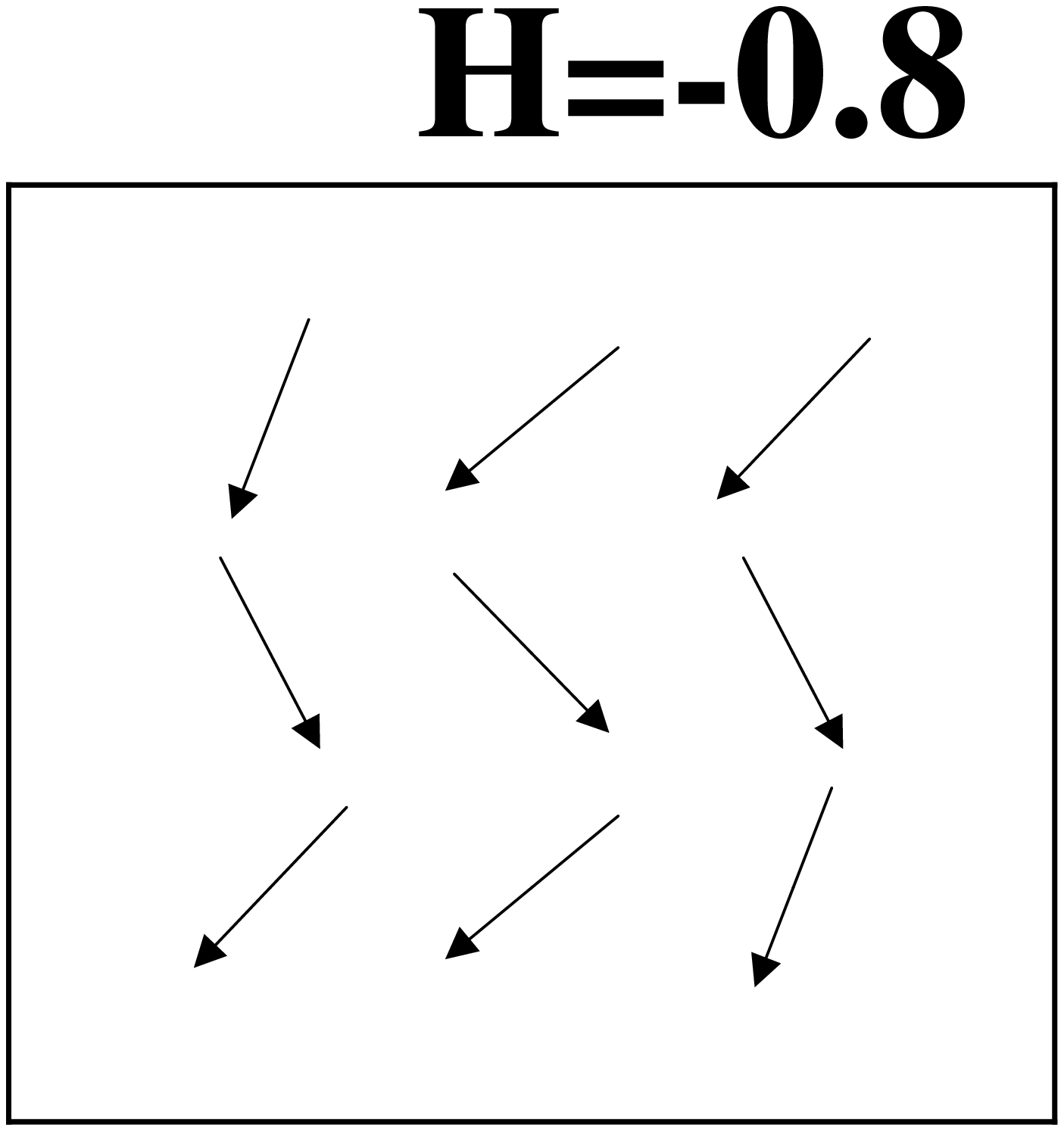}
  \hfill
  \includegraphics[angle=0,width=0.9in,totalheight=1.0in]{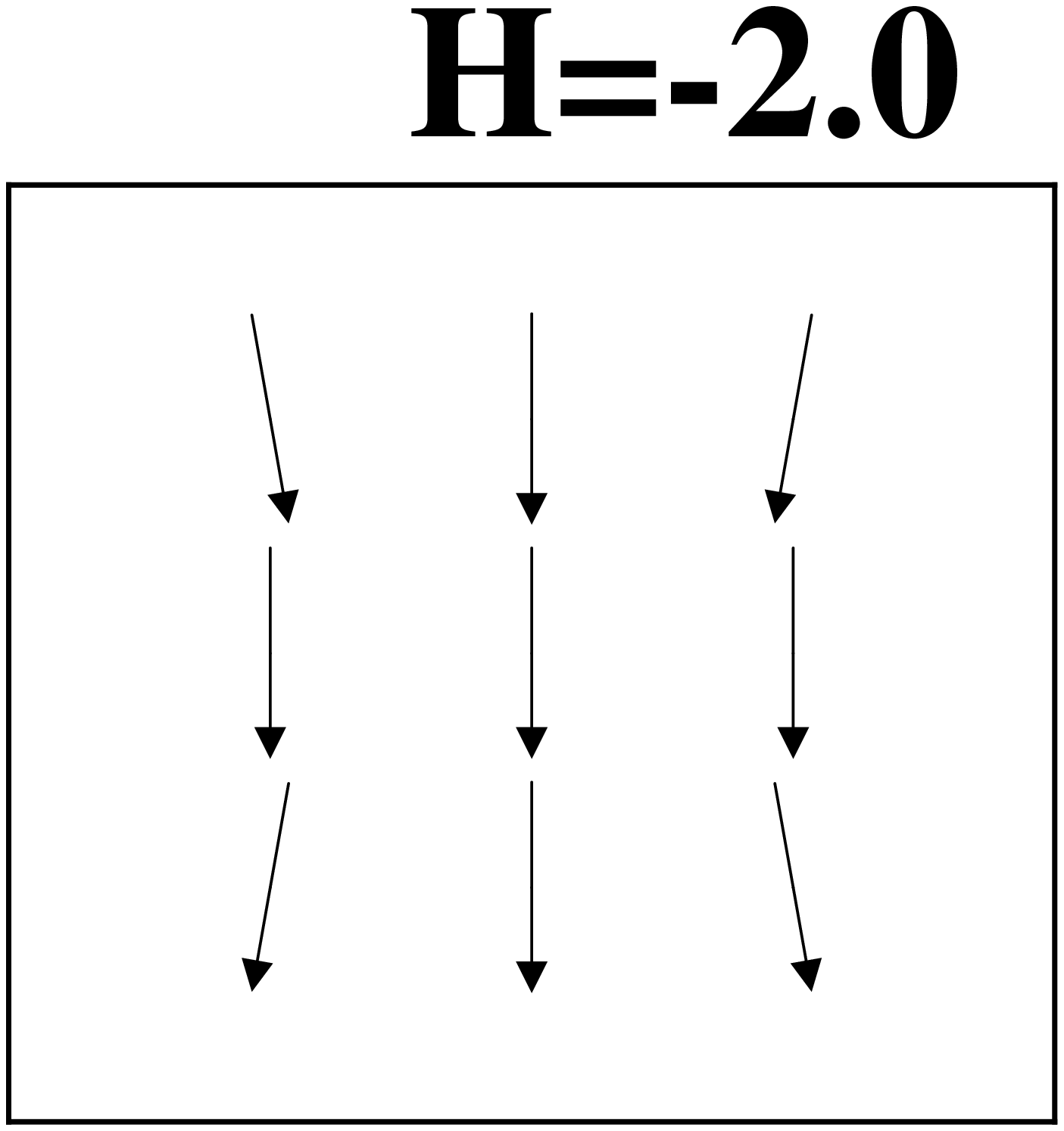}
  \caption{Spin arrangements for an array of $3\times 3$ ferromagnetic 
    nano dots in an external magnetic field.}
  \label{fig3}
\end{figure}

\begin{figure}[h]
  \centering
  \includegraphics[angle=0,width=0.9in,totalheight=1.0in]{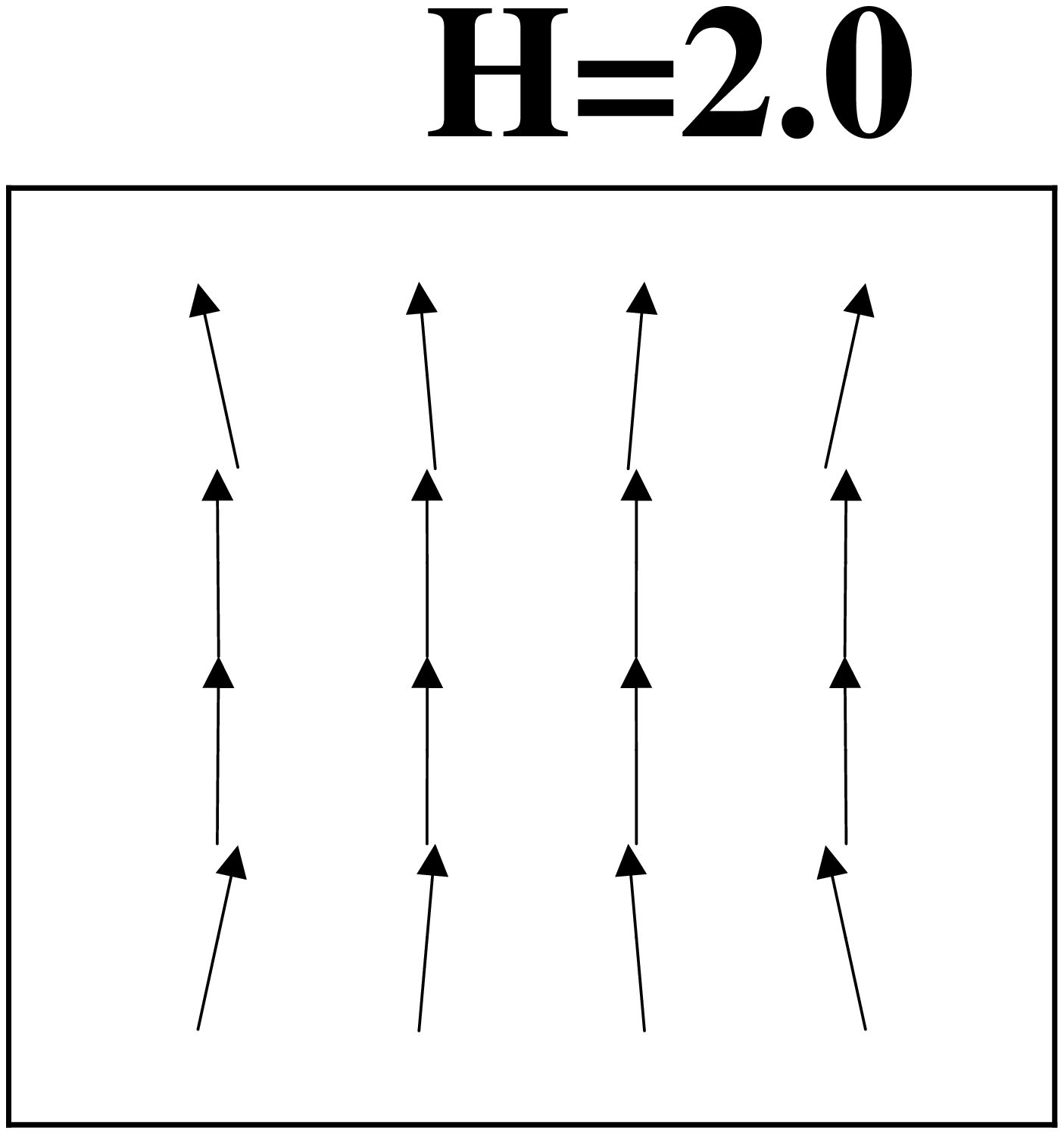}
  \hfill
  \includegraphics[angle=0,width=0.9in,totalheight=1.0in]{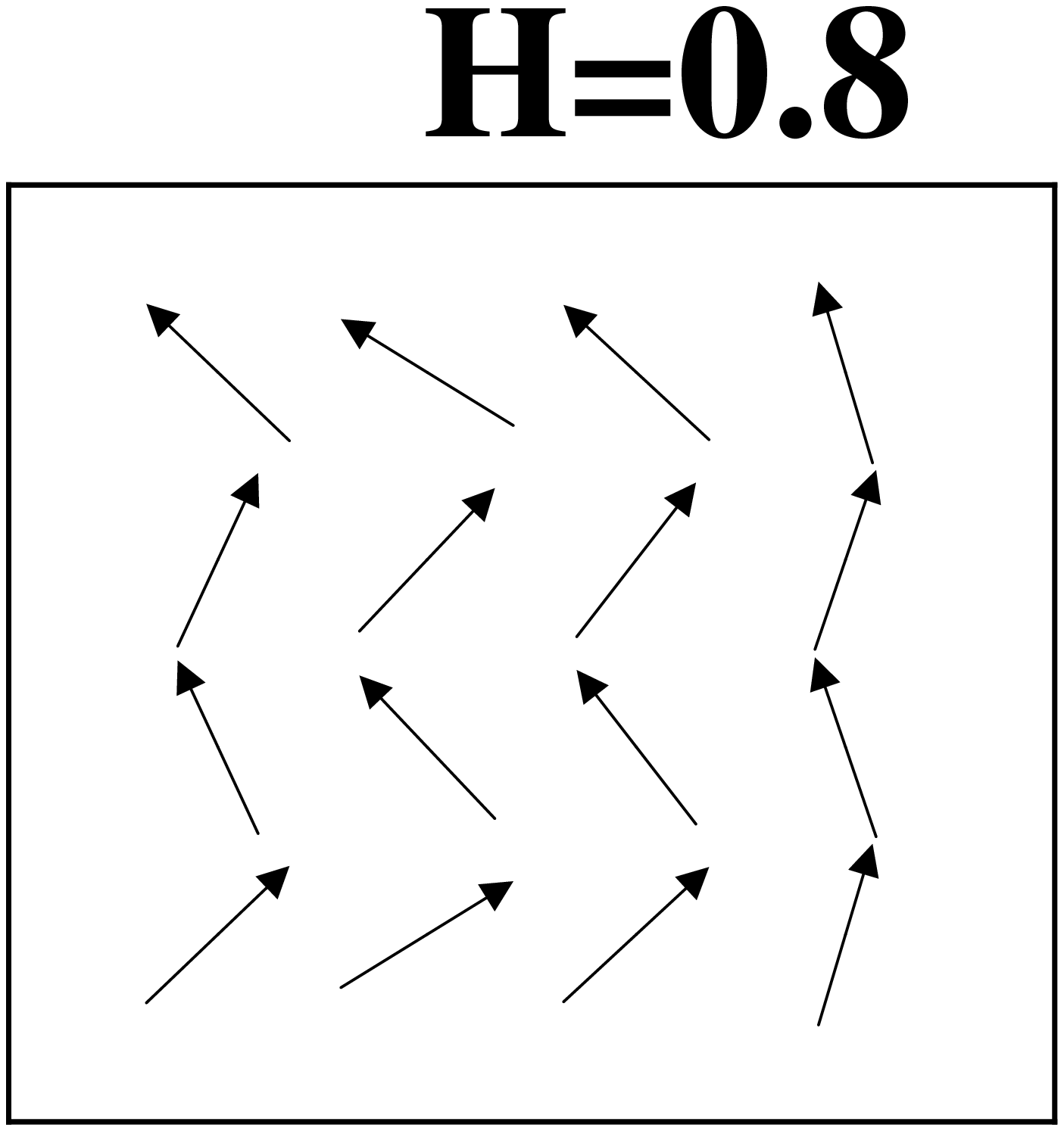}
  \hfill
  \includegraphics[angle=0,width=0.9in,totalheight=1.0in]{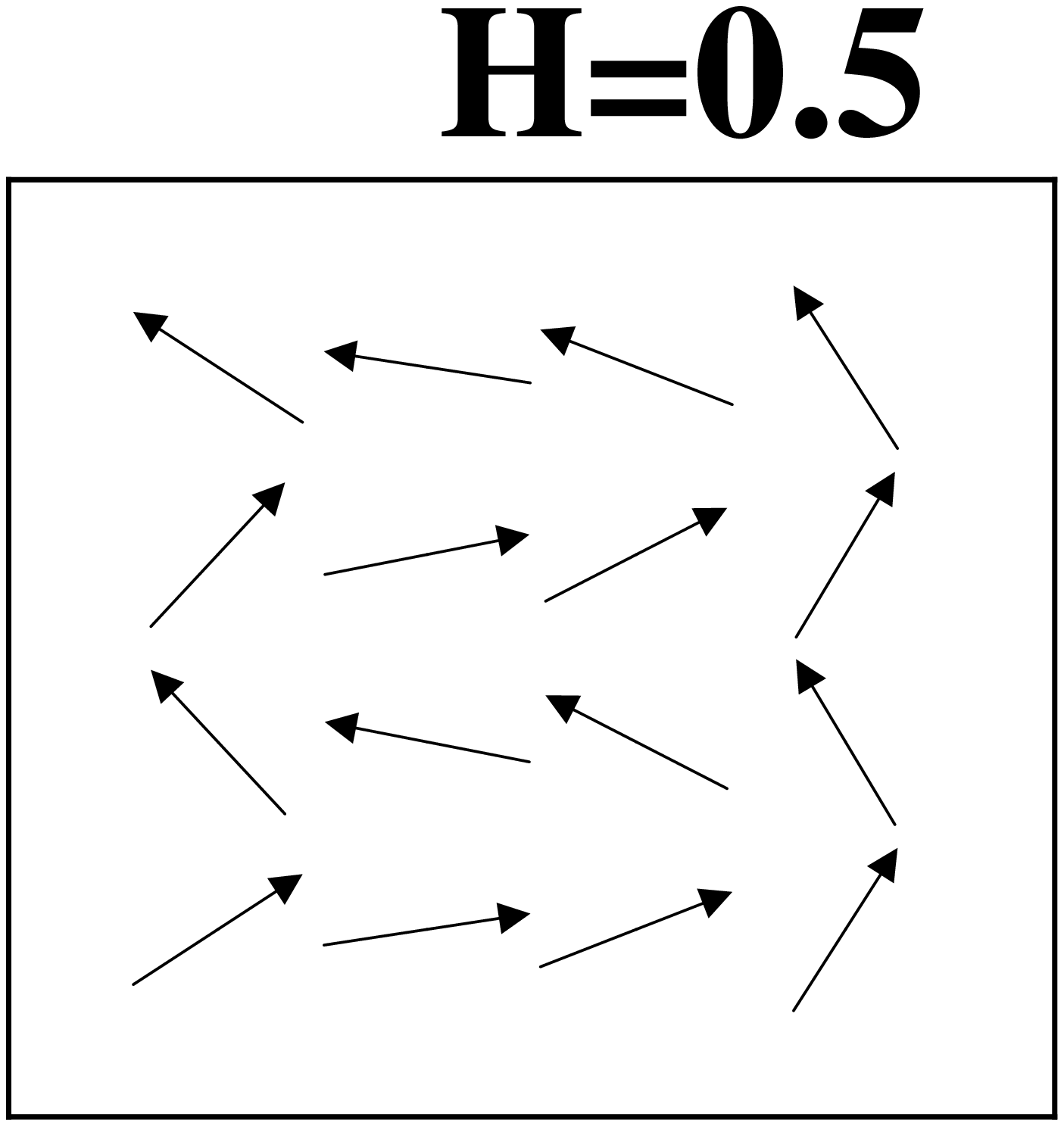} \\
  \vspace{0.15in}
  \includegraphics[angle=0,width=0.9in,totalheight=1.0in]{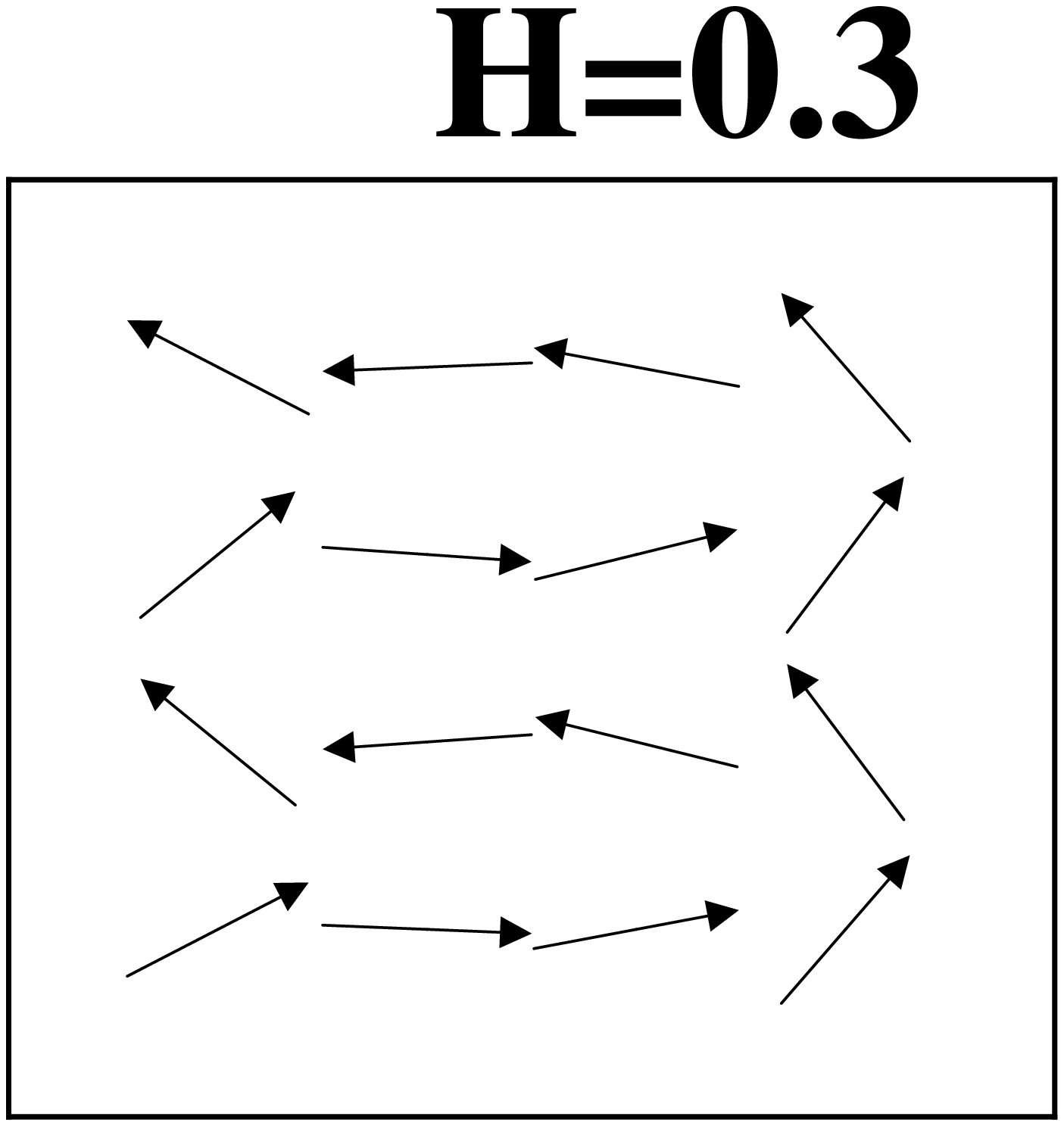}
  \hfill
  \includegraphics[angle=0,width=0.9in,totalheight=1.0in]{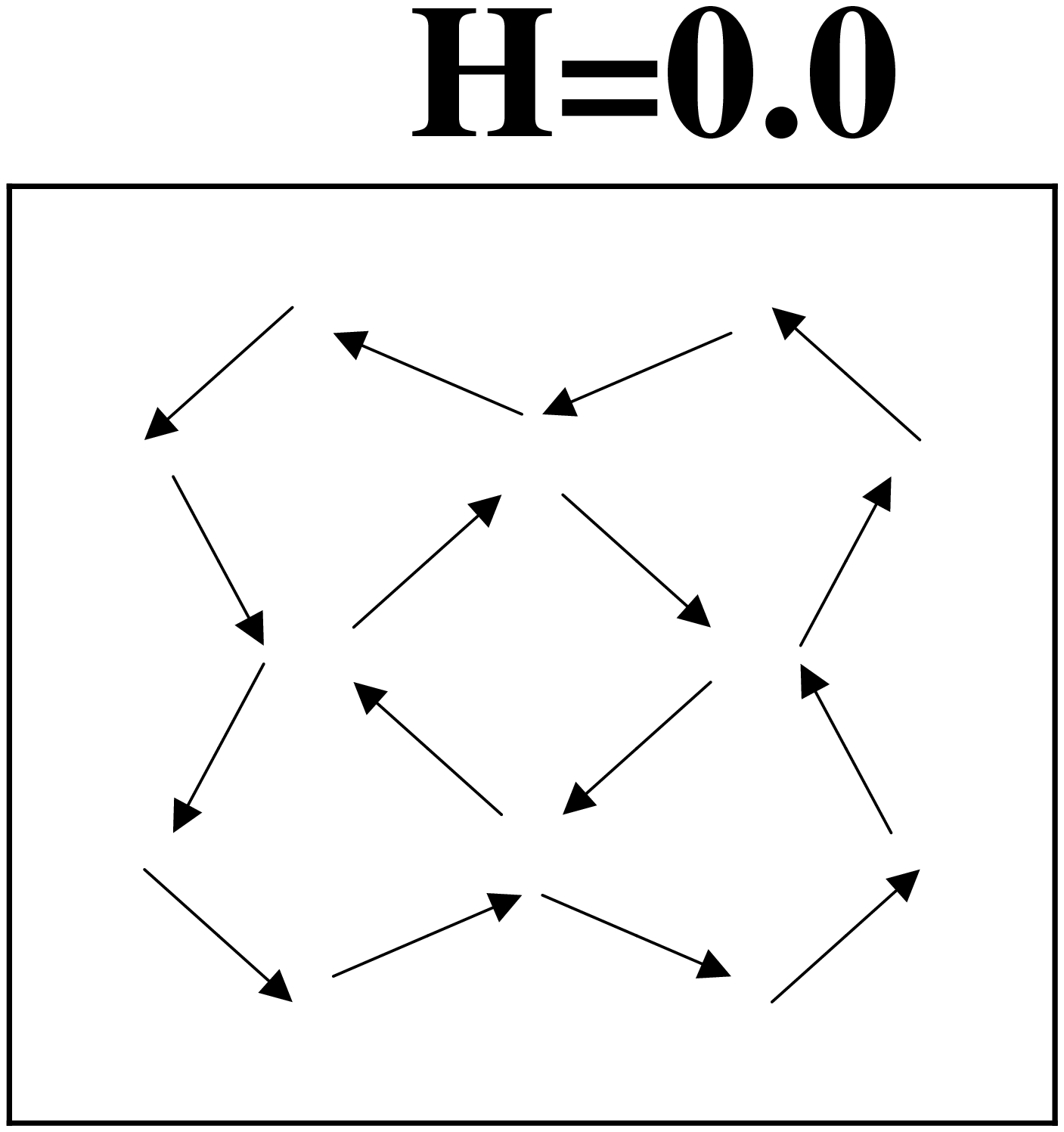}
  \hfill
  \includegraphics[angle=0,width=0.9in,totalheight=1.0in]{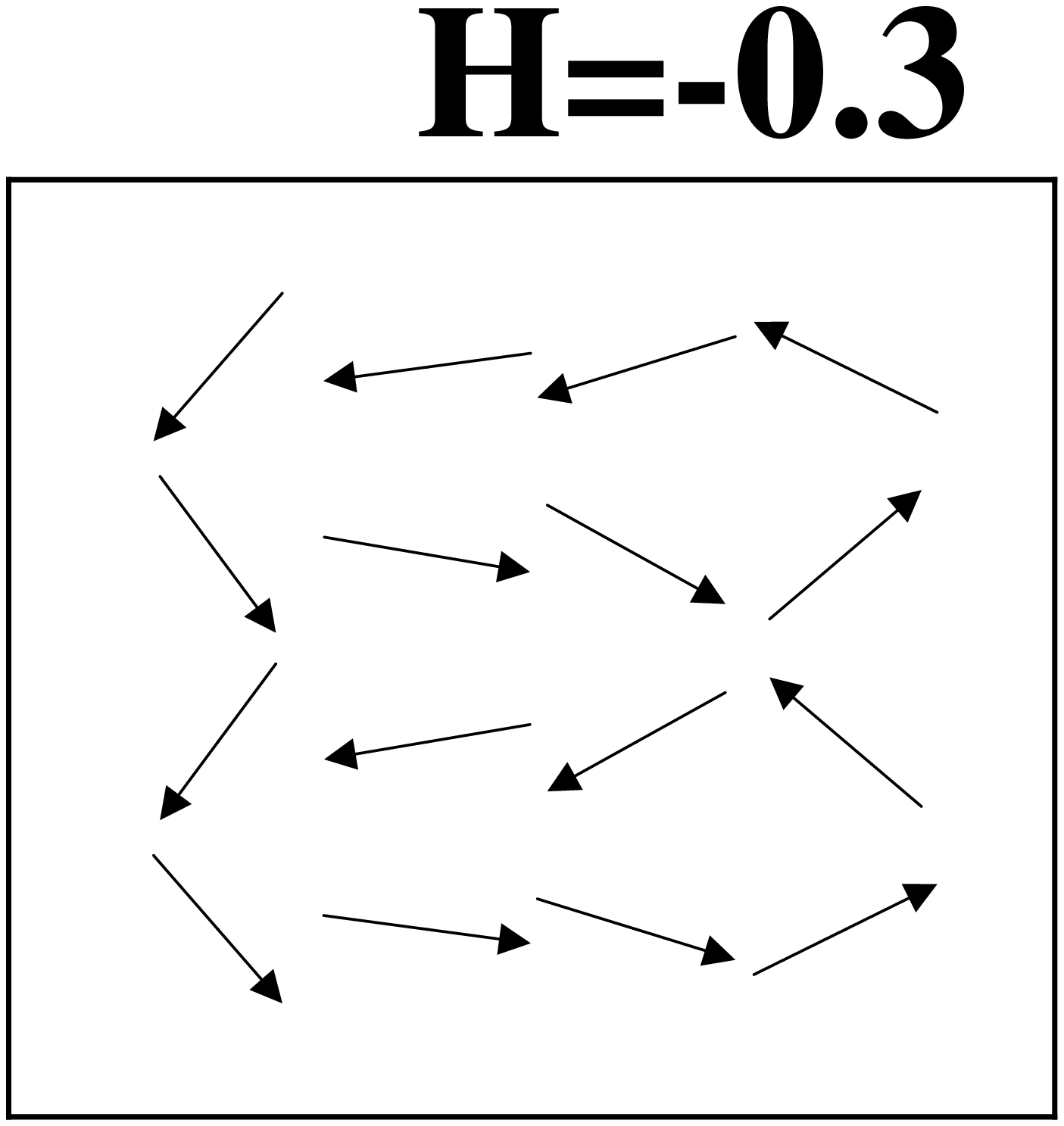}\\
  \vspace{0.15in}
  \includegraphics[angle=0,width=0.9in,totalheight=1.0in]{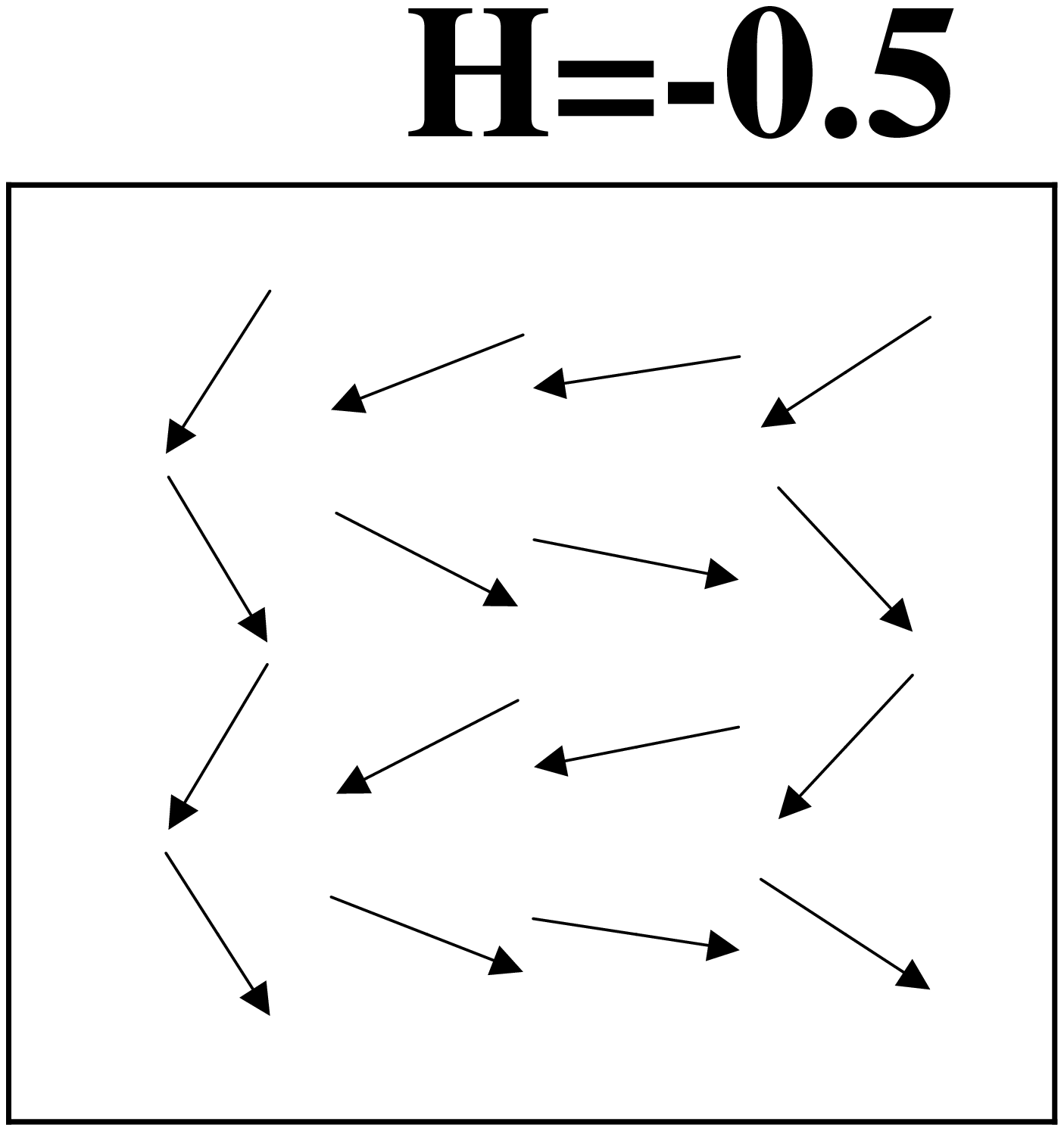}
  \hfill
  \includegraphics[angle=0,width=0.9in,totalheight=1.0in]{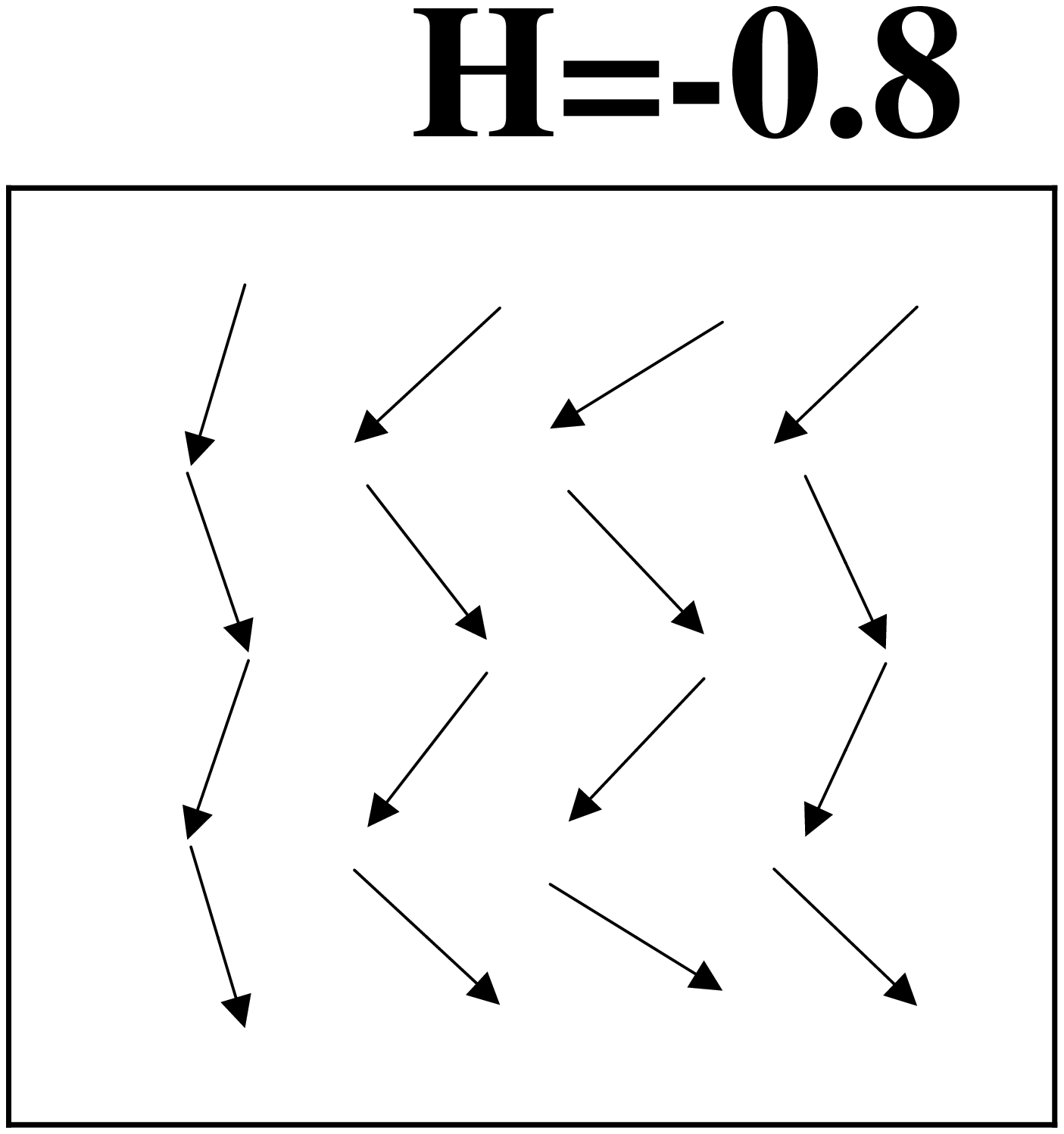}
  \hfill
  \includegraphics[angle=0,width=0.9in,totalheight=1.0in]{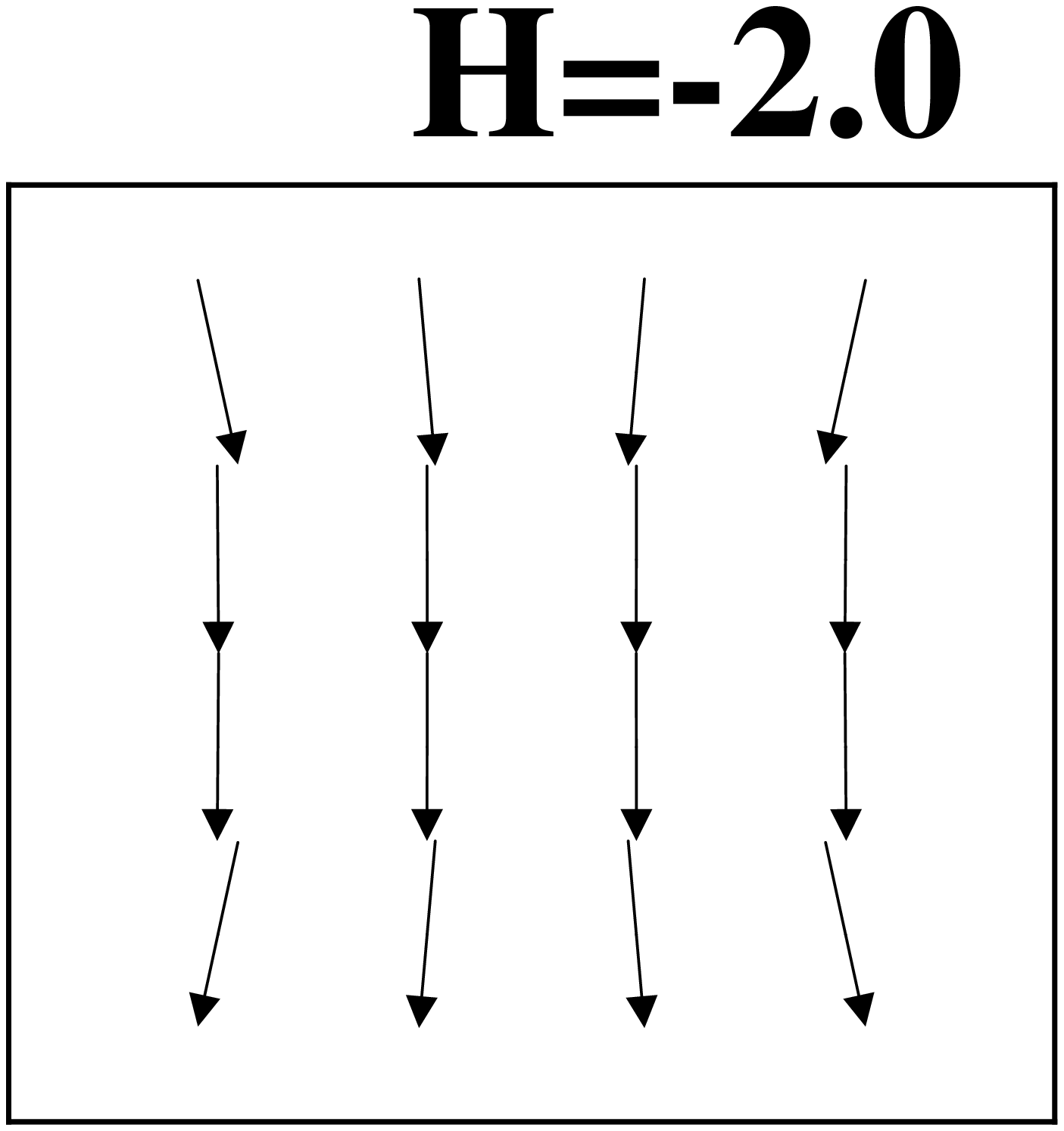}
  \caption{Spin arrangements for an array of $4\times 4$ ferromagnetic 
    nano dots in external magnetic field.}
  \label{fig4}
\end{figure}

\begin{figure}[h]
  \centering
  \includegraphics[angle=0,width=0.9in,totalheight=1.0in]{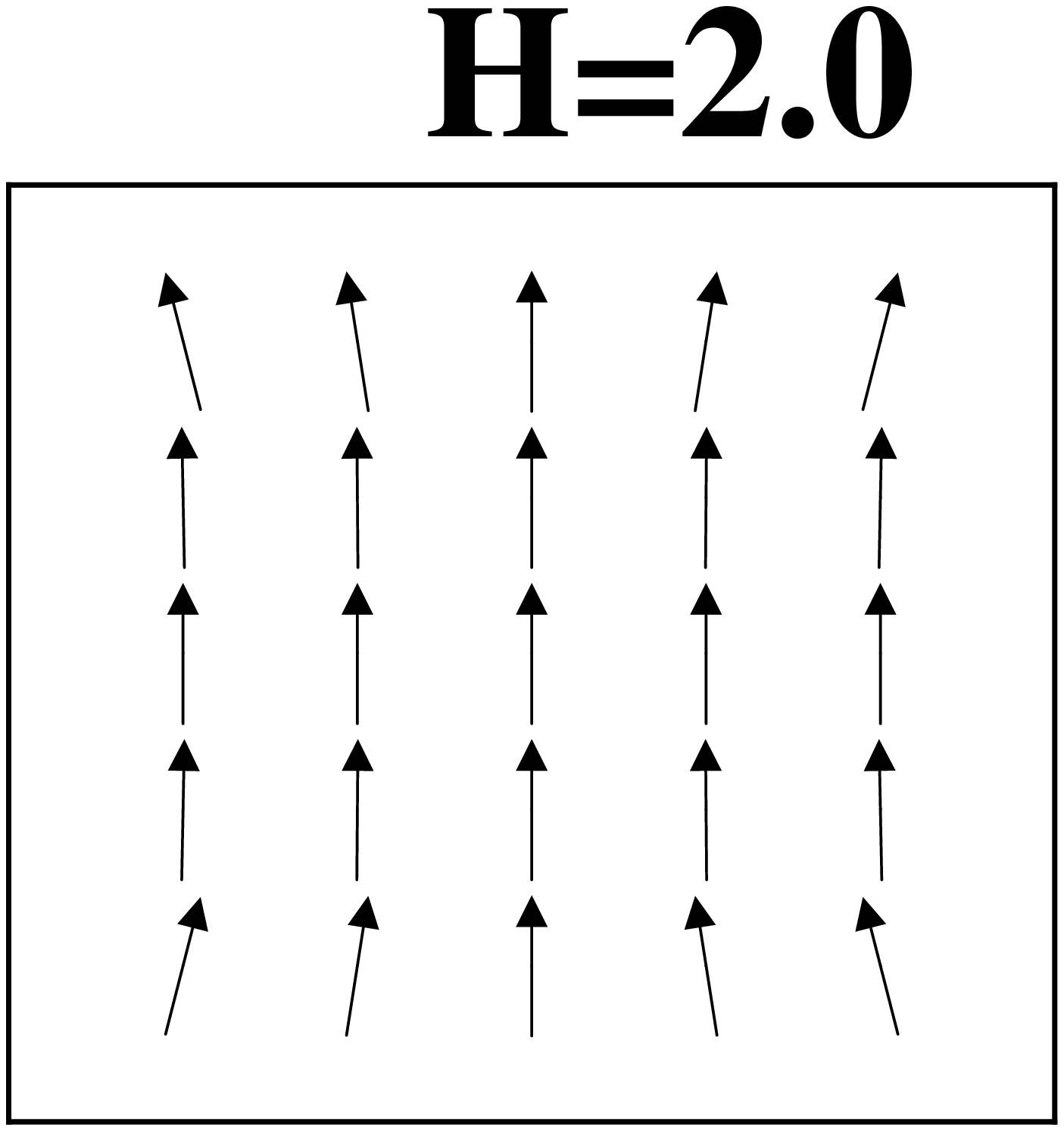}
  \hfill
  \includegraphics[angle=0,width=0.9in,totalheight=1.0in]{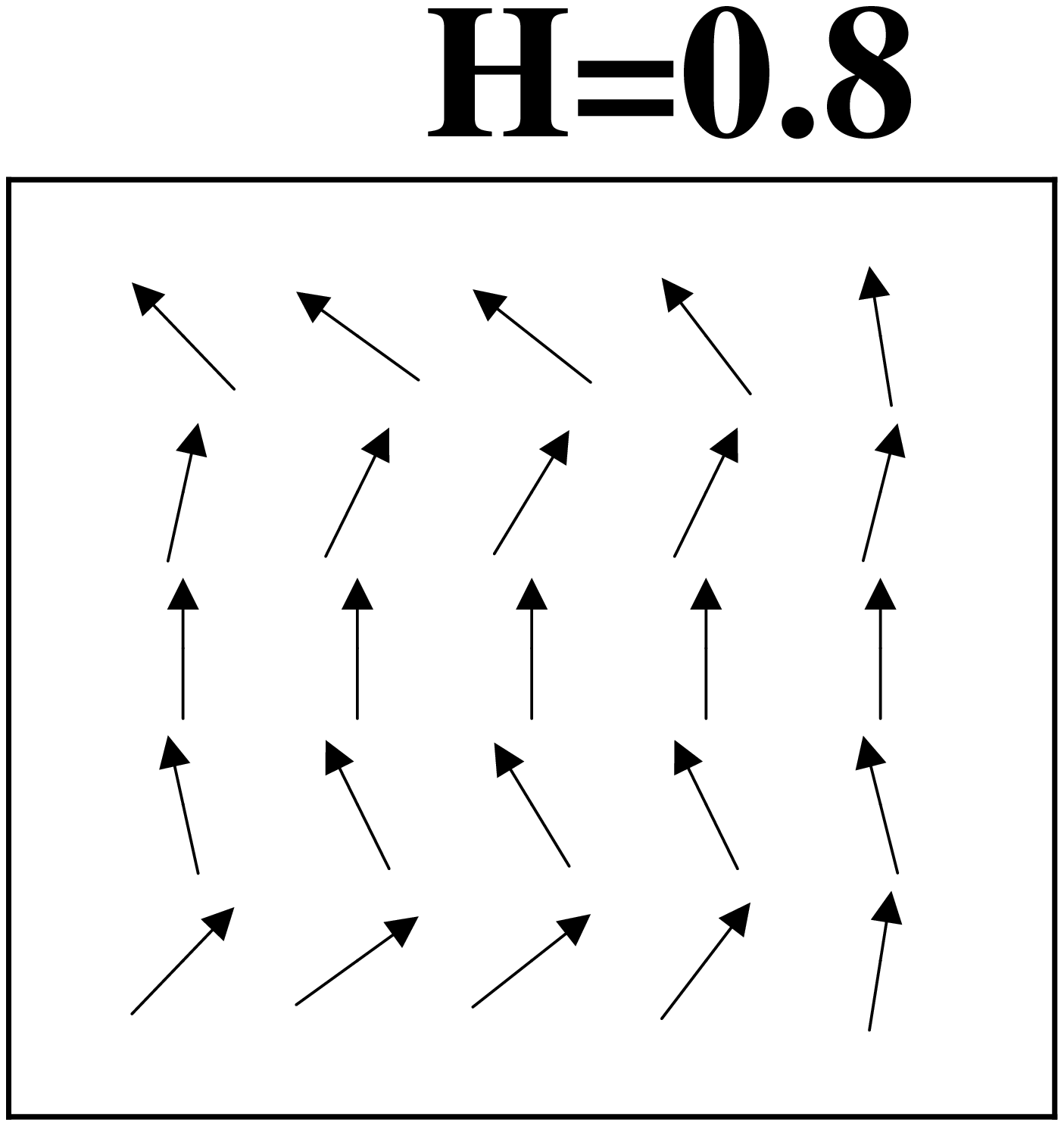}
  \hfill
  \includegraphics[angle=0,width=0.9in,totalheight=1.0in]{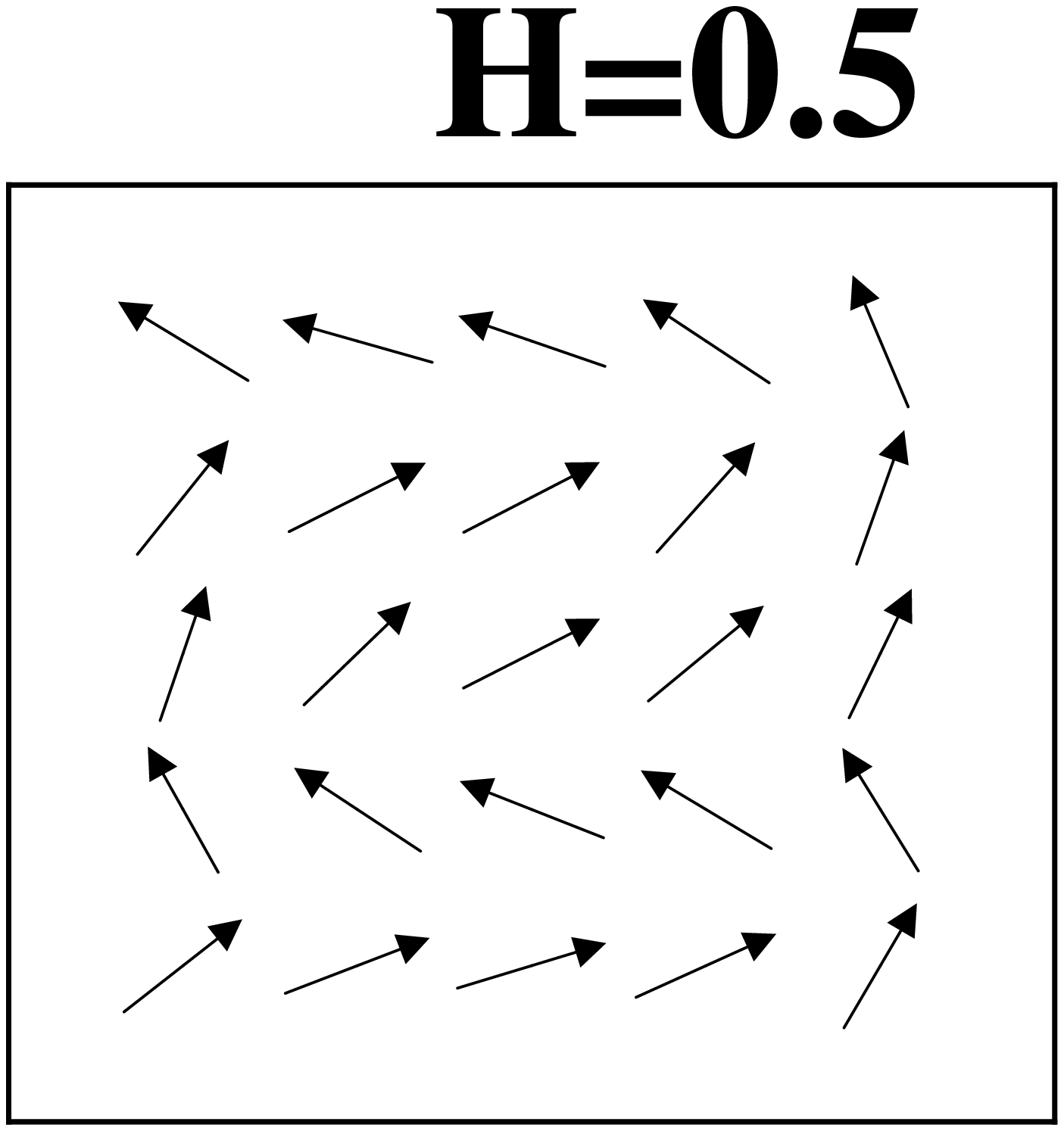} \\
  \vspace{0.15in}
  \includegraphics[angle=0,width=0.9in,totalheight=1.0in]{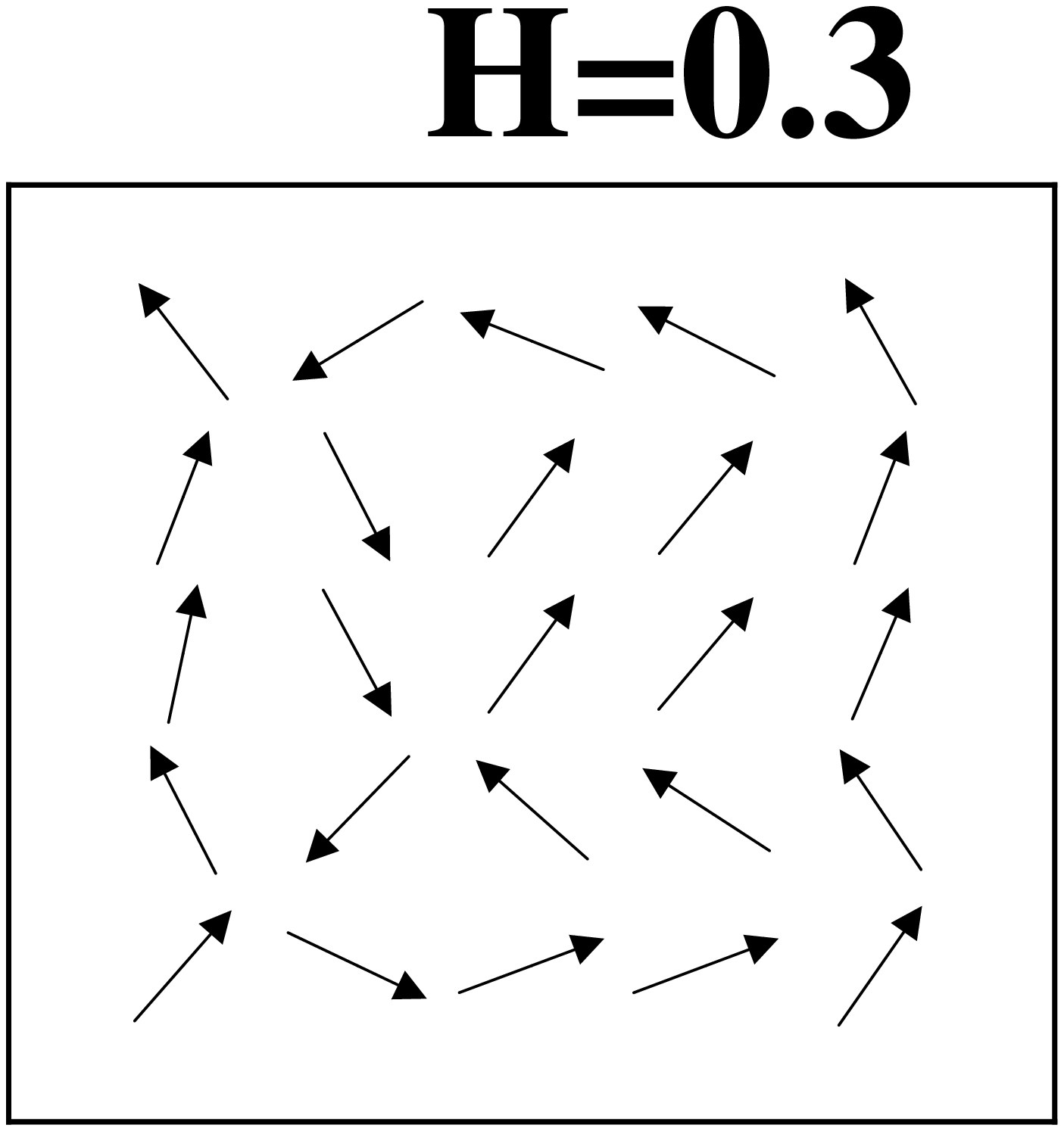}
  \hfill
  \includegraphics[angle=0,width=0.9in,totalheight=1.0in]{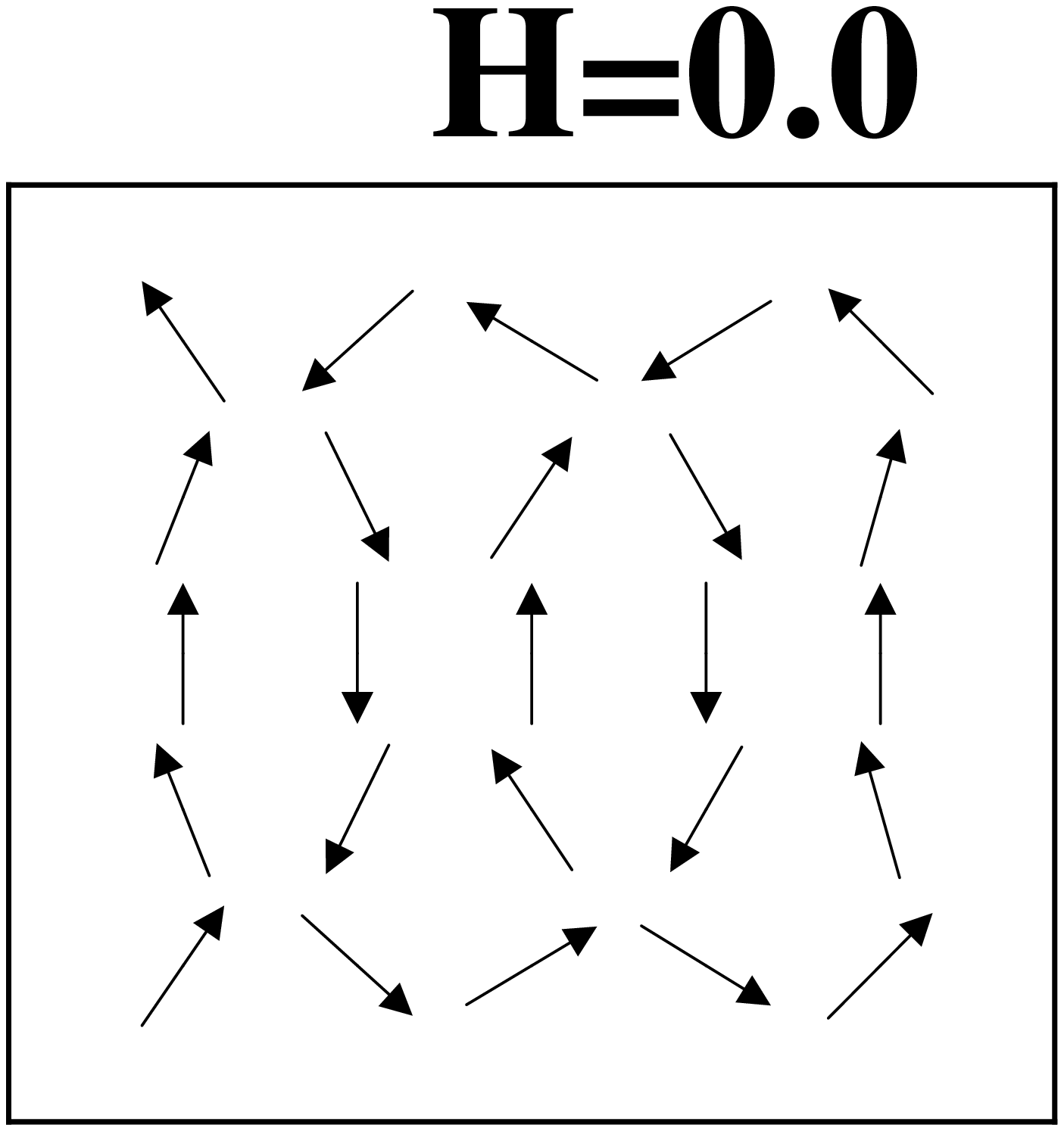}
  \hfill
  \includegraphics[angle=0,width=0.9in,totalheight=1.0in]{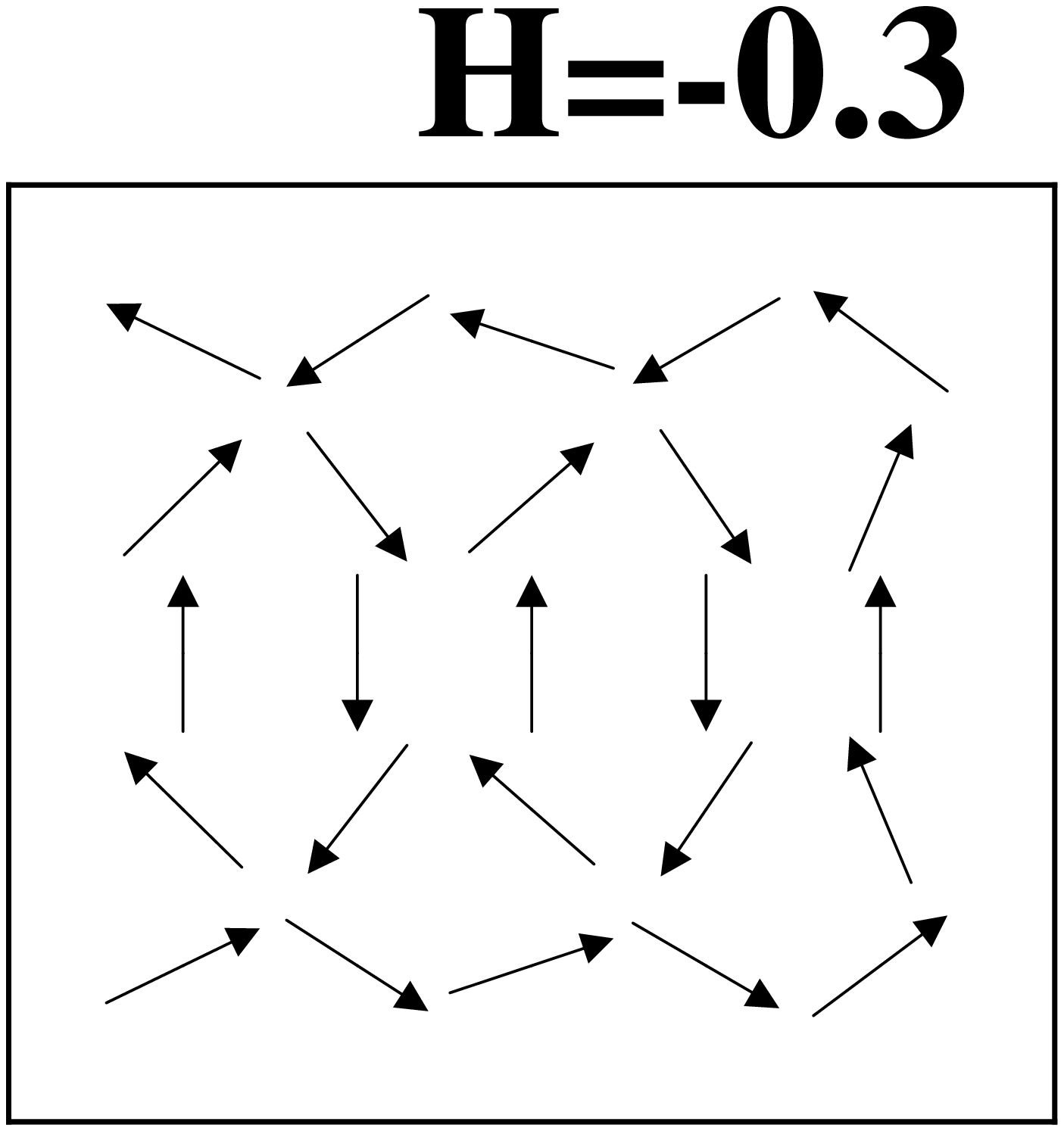} \\
  \vspace{0.15in}
  \includegraphics[angle=0,width=0.9in,totalheight=1.0in]{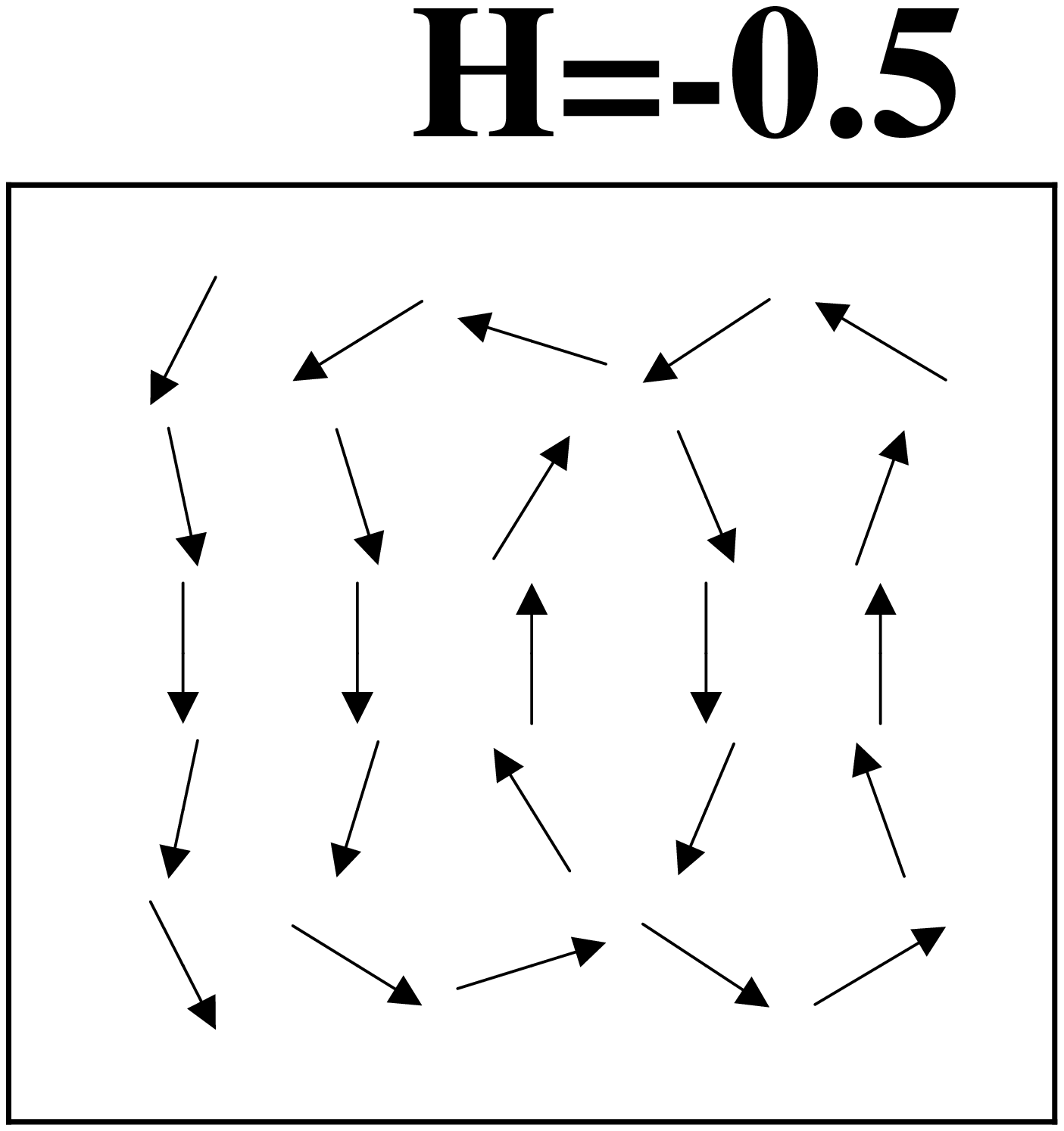}
  \hfill
  \includegraphics[angle=0,width=0.9in,totalheight=1.0in]{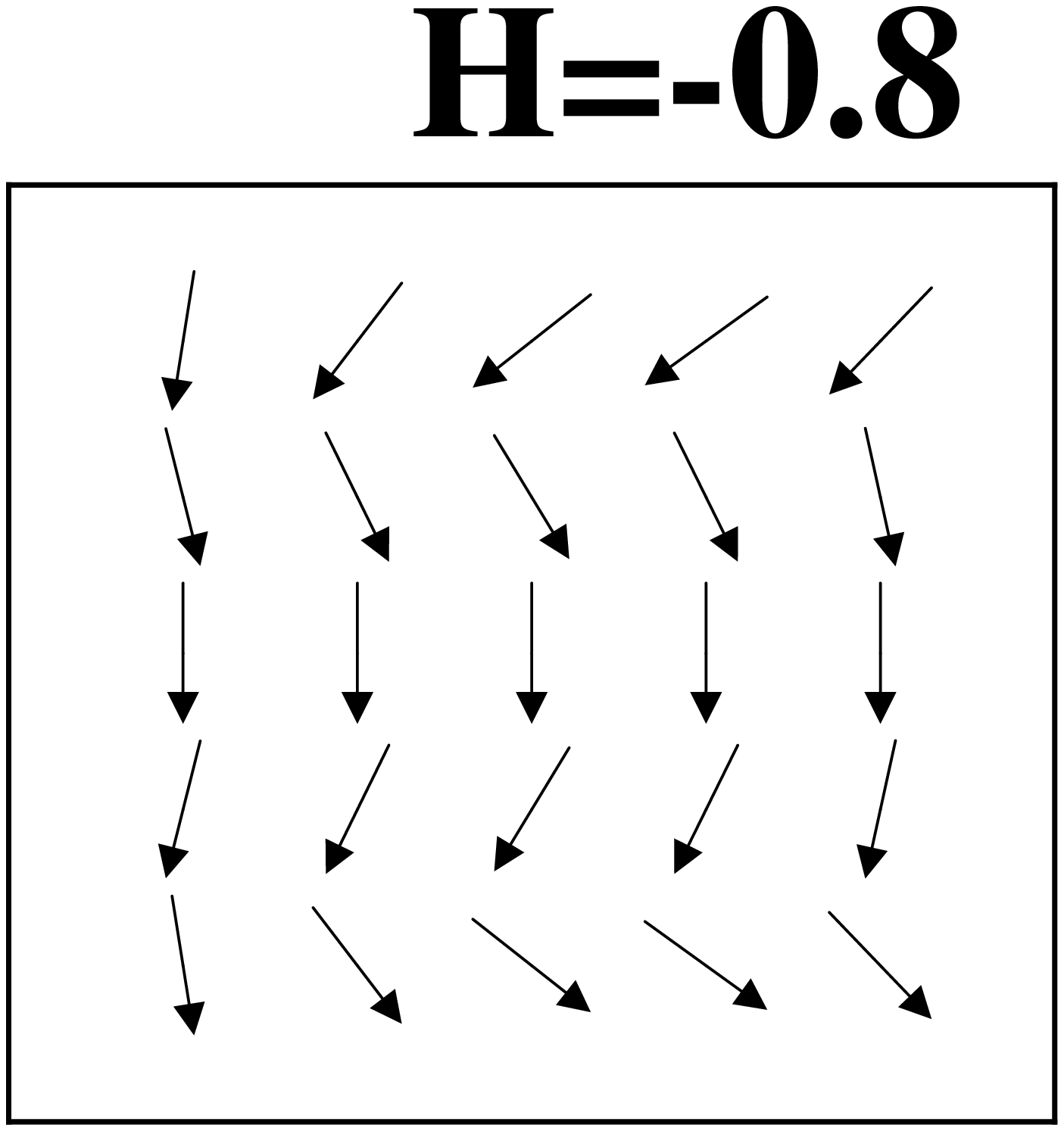}
  \hfill
  \includegraphics[angle=0,width=0.9in,totalheight=1.0in]{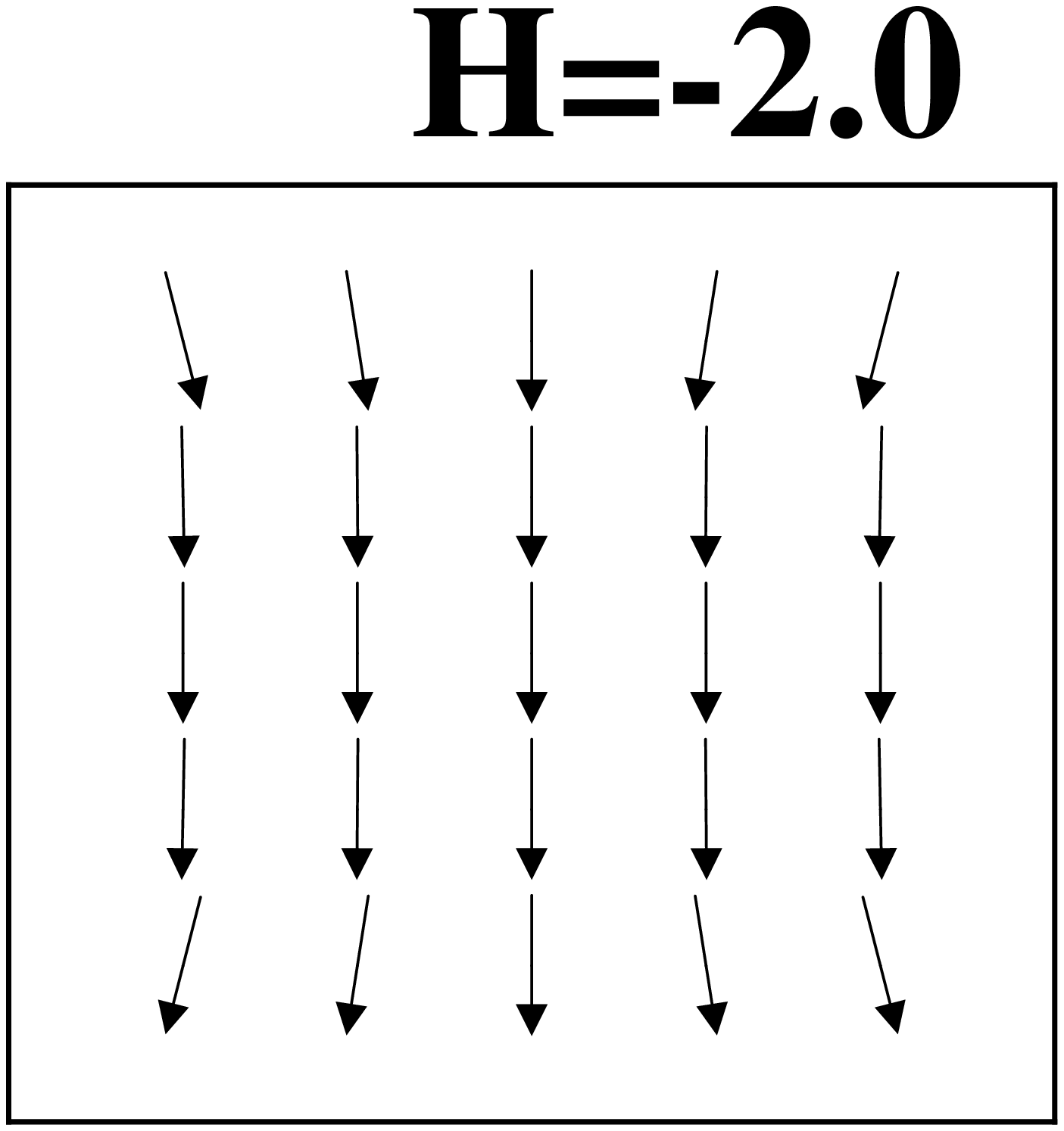}
  \caption{Spin arrangements for an array of $5\times 5$ ferromagnetic 
    nano dots in external magnetic field.}
  \label{fig5}
\end{figure}

In the absence of an external magnetic field the array of
$N\times N$ nano-dots favors antiferromagnetic ordering, thus
minimizing its magnetostatic energy. A large external magnetic field
applied to the array tends to orient the magnetic moments 
along the field, thus minimizing the Zeeman field energy. 
However, the spins at the array corners tip by a small angle, as shown 
in Fig.~\ref{fig2}-Fig.~\ref{fig5}, forming a two-dimensional
``flower`` state \cite{aharoni}, \cite{TLW}. The flower state persists
until the applied field falls to $H_0 =M_s$. For lower values of the
applied field, the competition between the dipole-dipole interaction and
the Zeeman energy becomes significant, and changes the array ordering.

\begin{figure}[h]
  \centering
  \includegraphics[angle=0,width=0.9in,totalheight=1.0in]{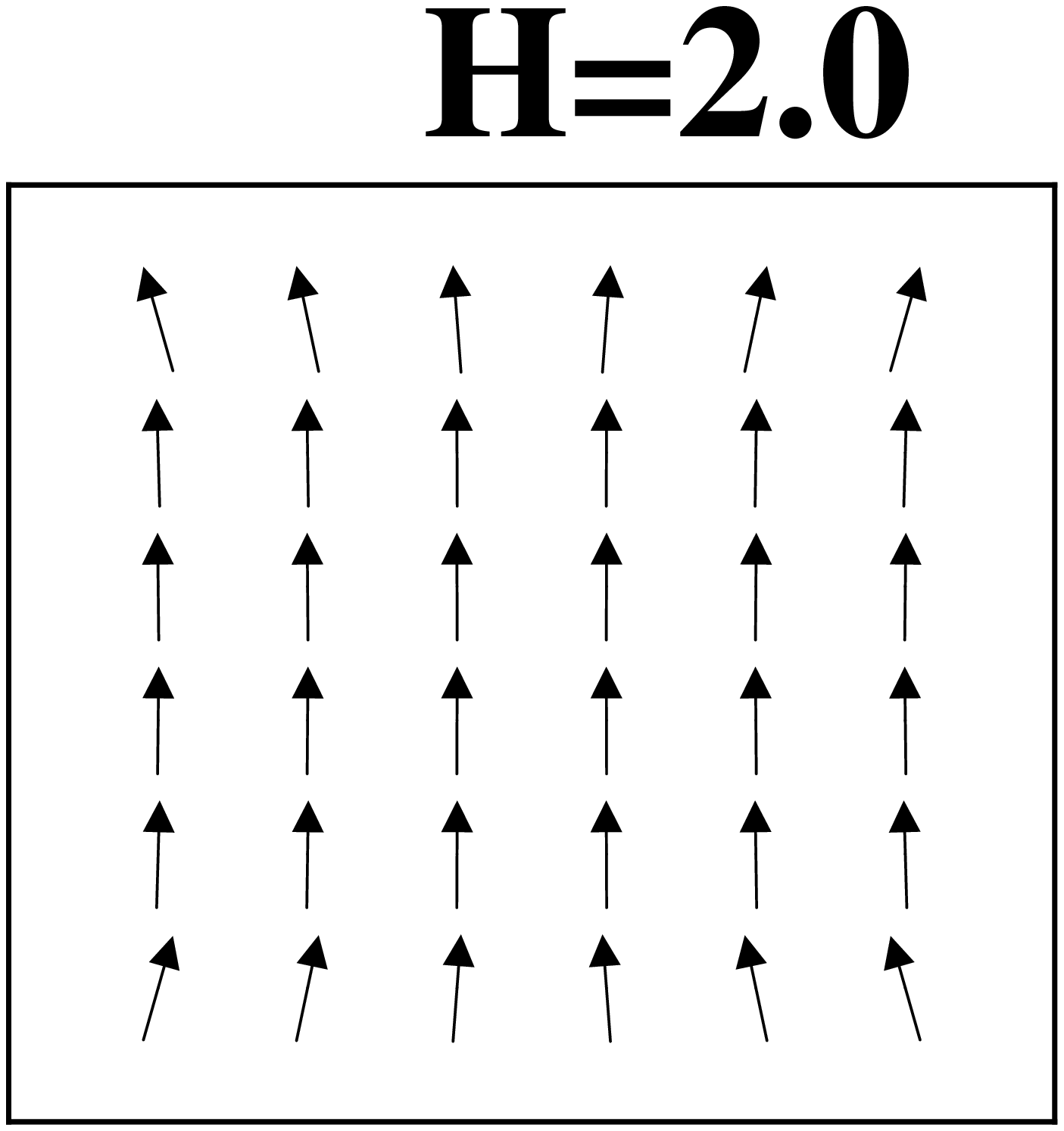}
  \hfill
  \includegraphics[angle=0,width=0.9in,totalheight=1.0in]{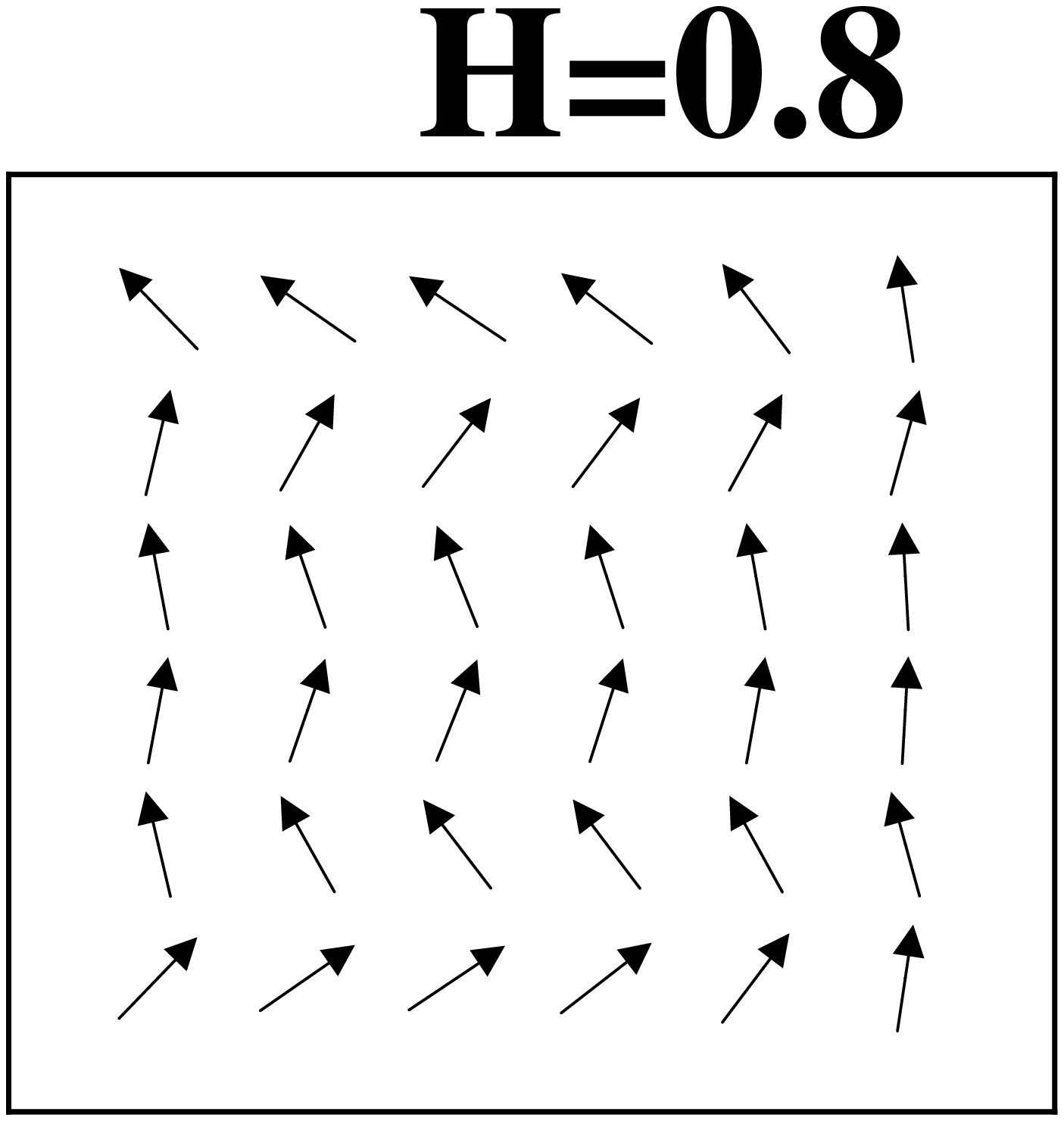}
  \hfill
  \includegraphics[angle=0,width=0.9in,totalheight=1.0in]{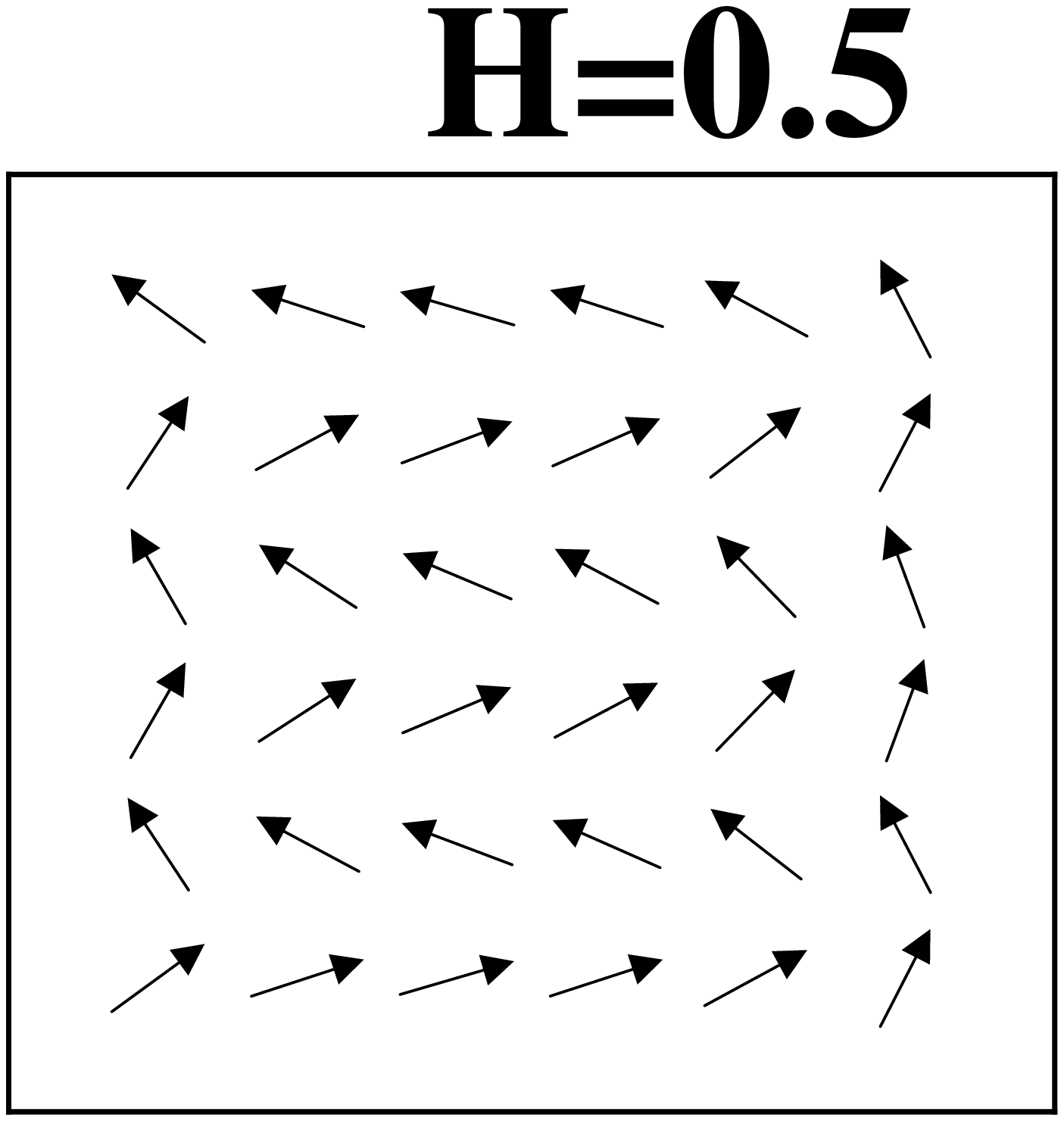}\\
  \vspace{0.15in}
  \includegraphics[angle=0,width=0.9in,totalheight=1.0in]{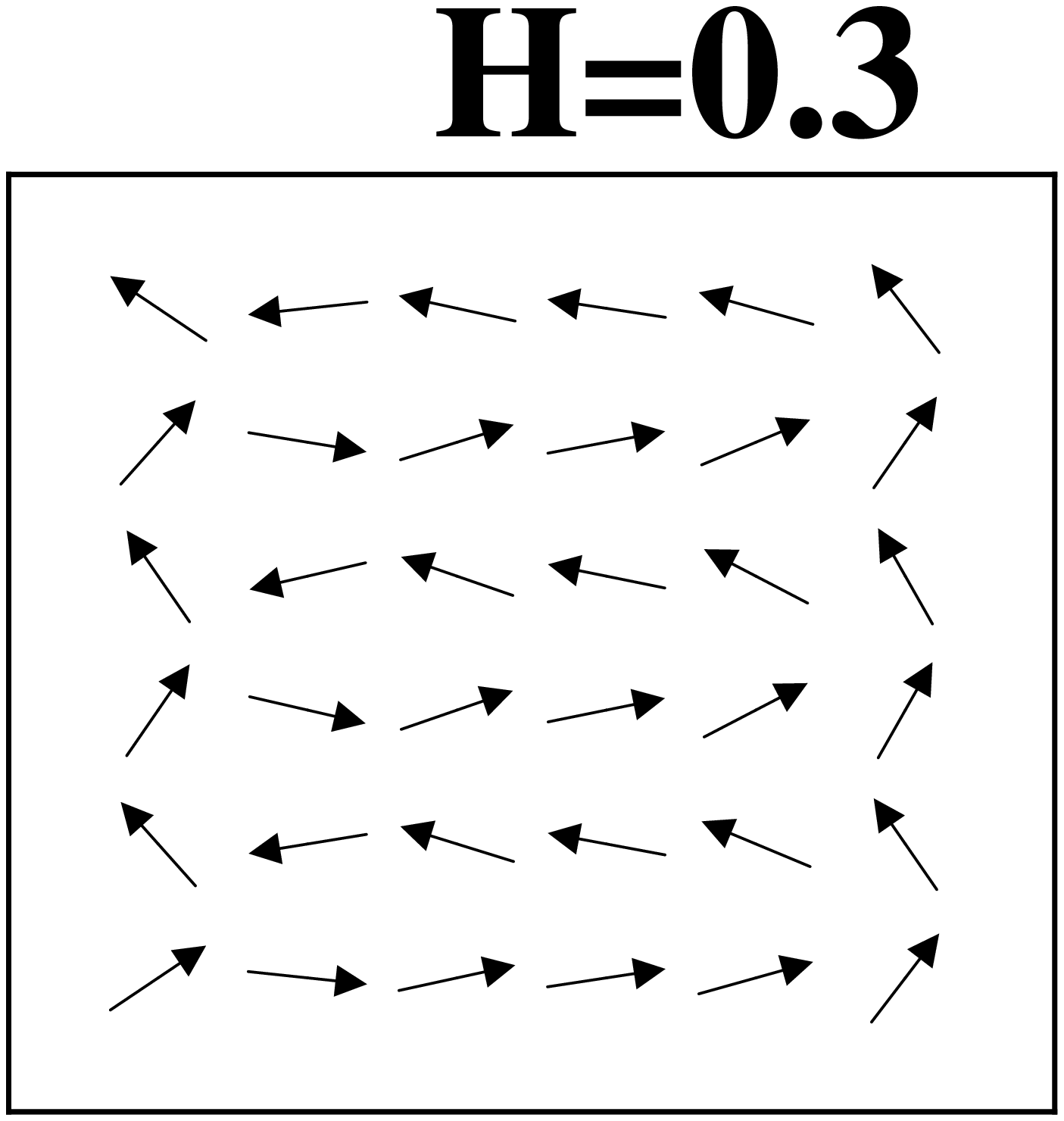}
  \hfill
  \includegraphics[angle=0,width=0.9in,totalheight=1.0in]{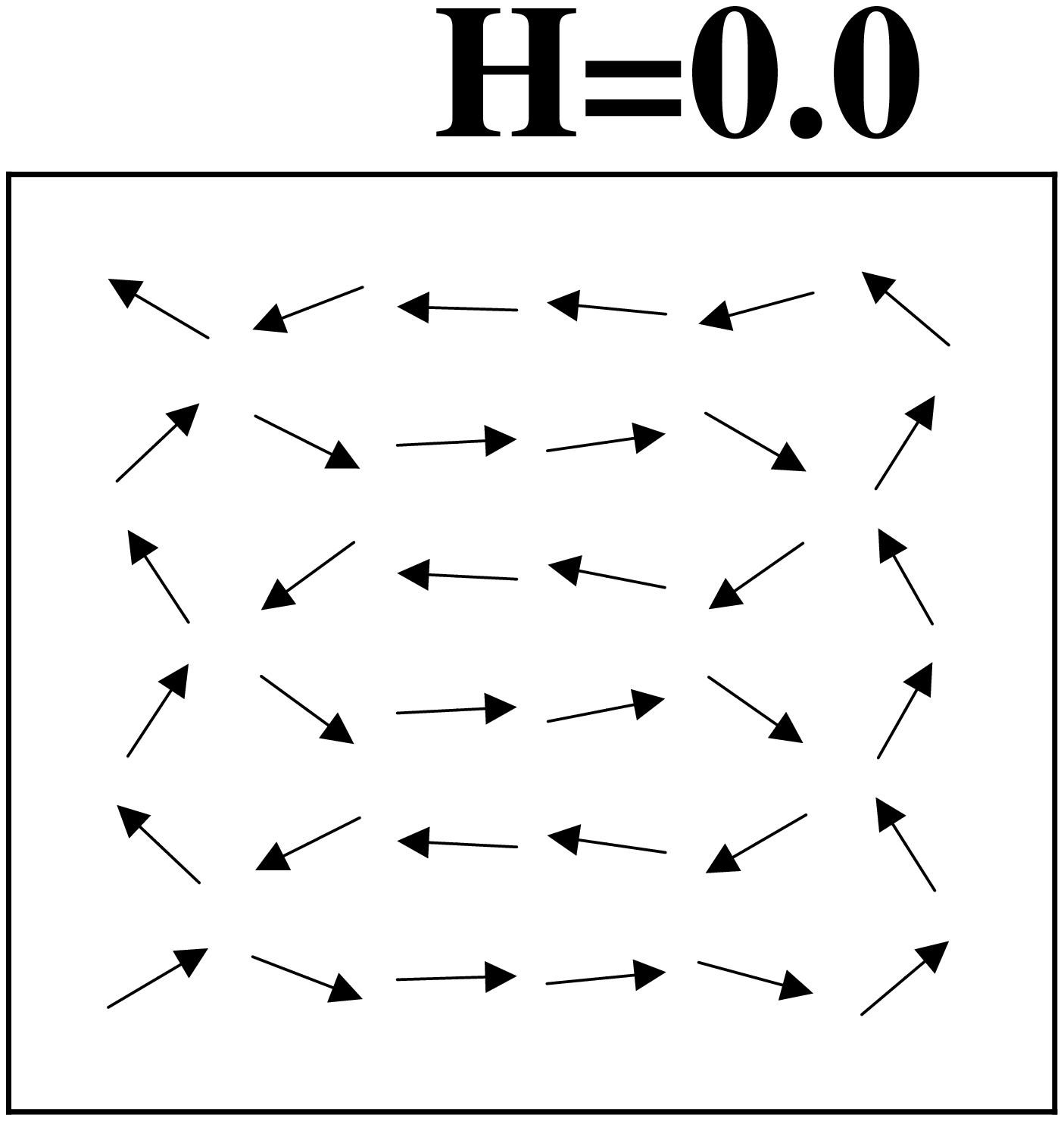}
  \hfill
  \includegraphics[angle=0,width=0.9in,totalheight=1.0in]{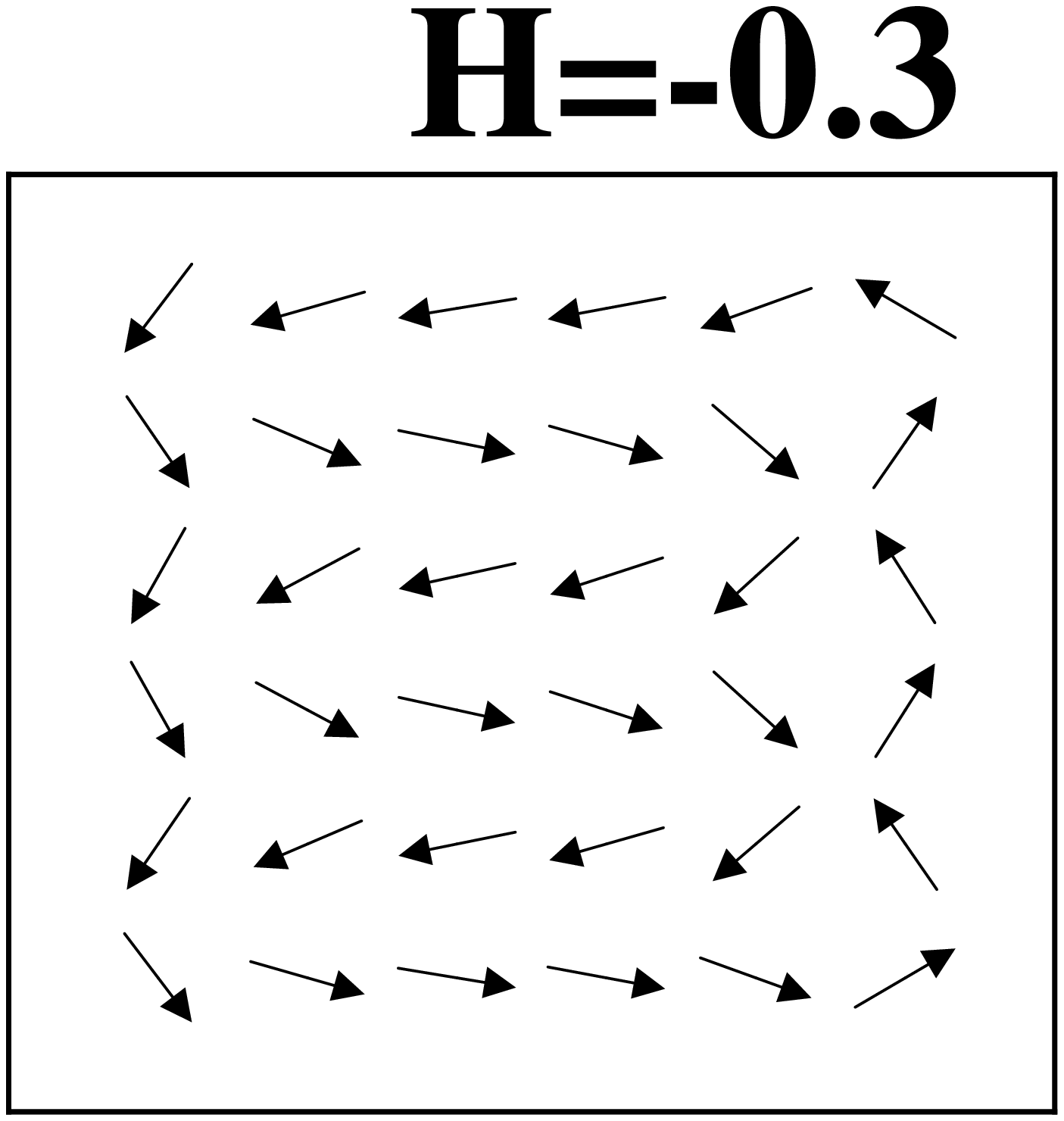}\\
  \vspace{0.15in}
  \includegraphics[angle=0,width=0.9in,totalheight=1.0in]{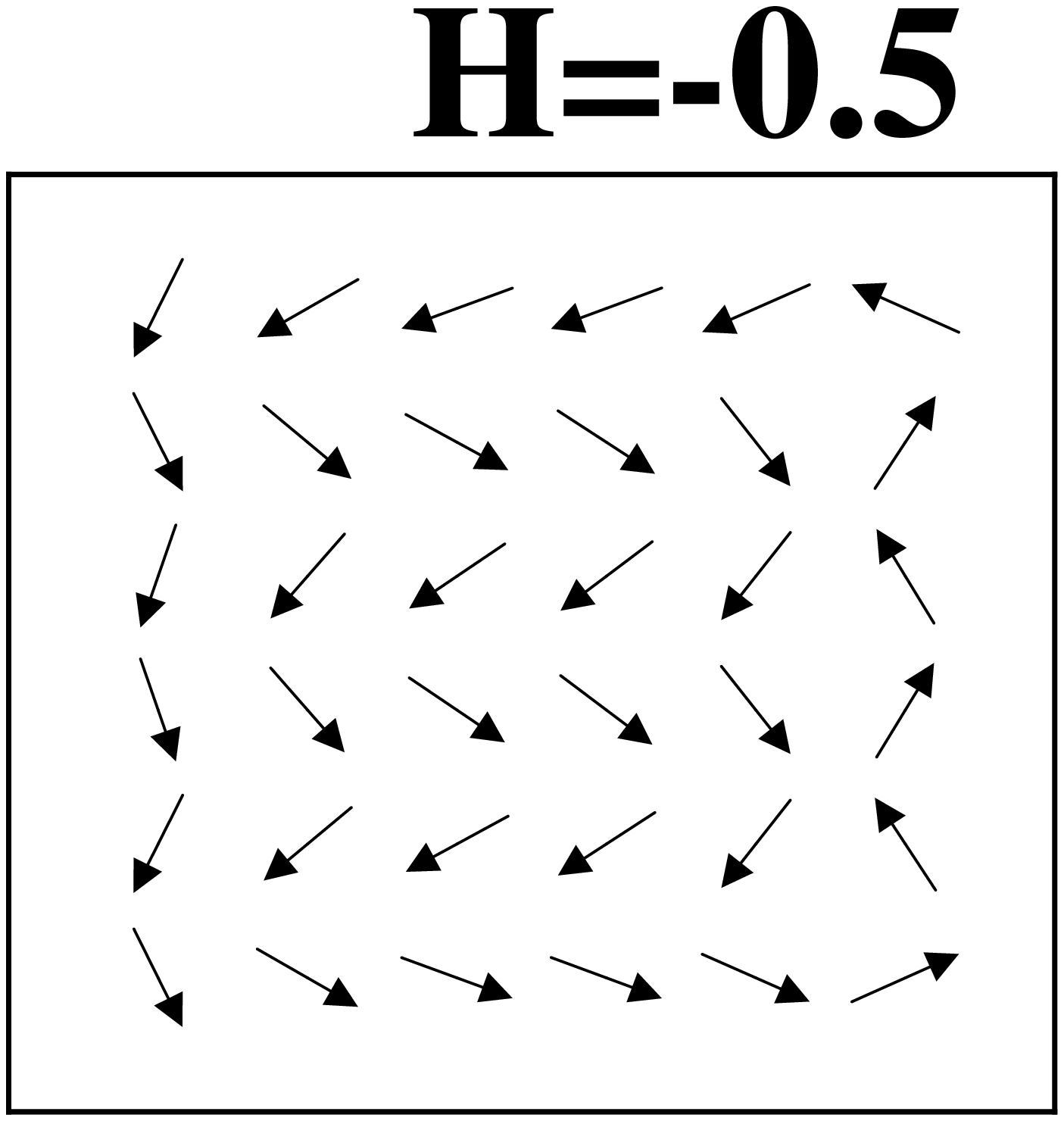}
  \hfill
  \includegraphics[angle=0,width=0.9in,totalheight=1.0in]{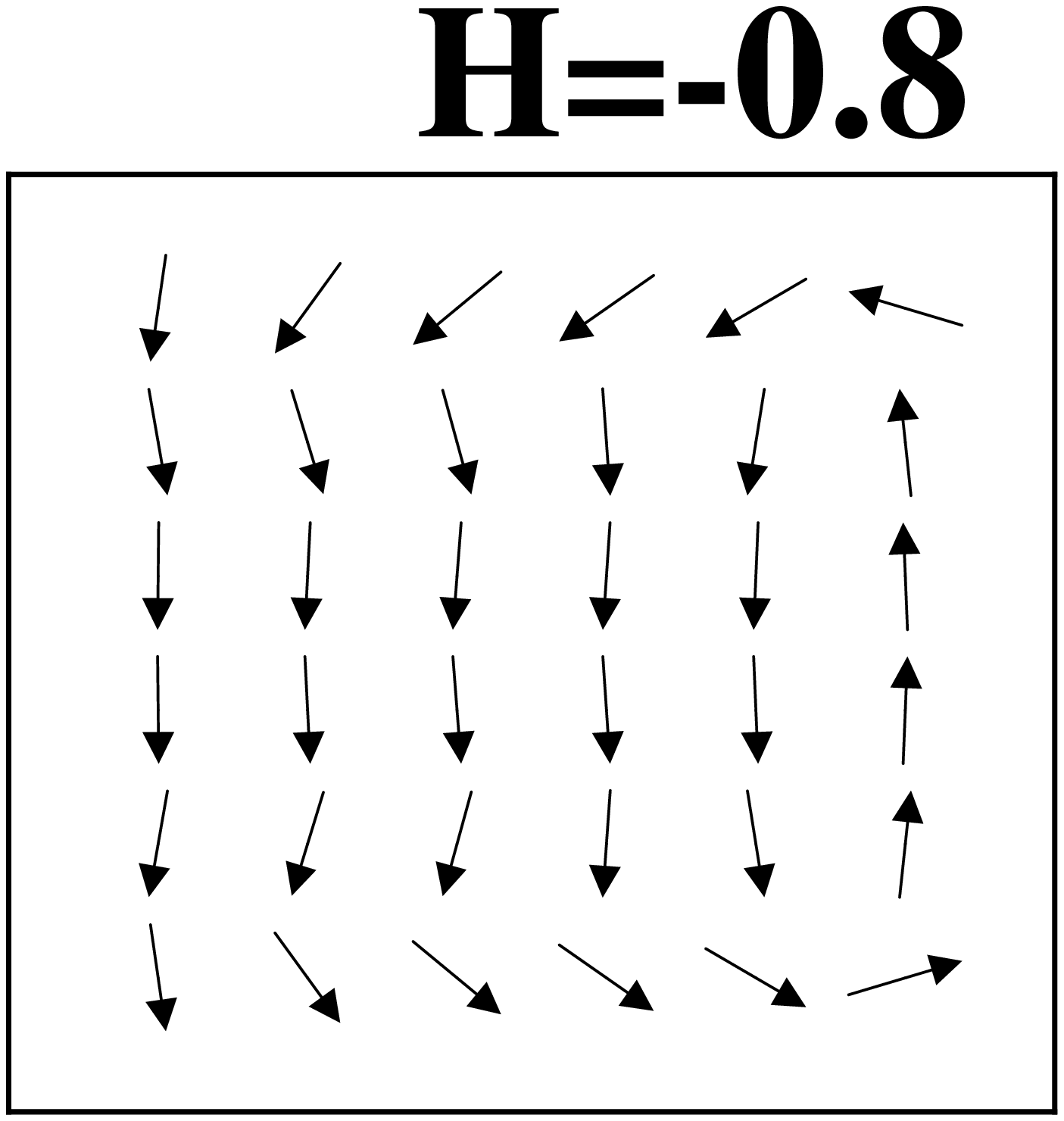}
  \hfill
  \includegraphics[angle=0,width=0.9in,totalheight=1.0in]{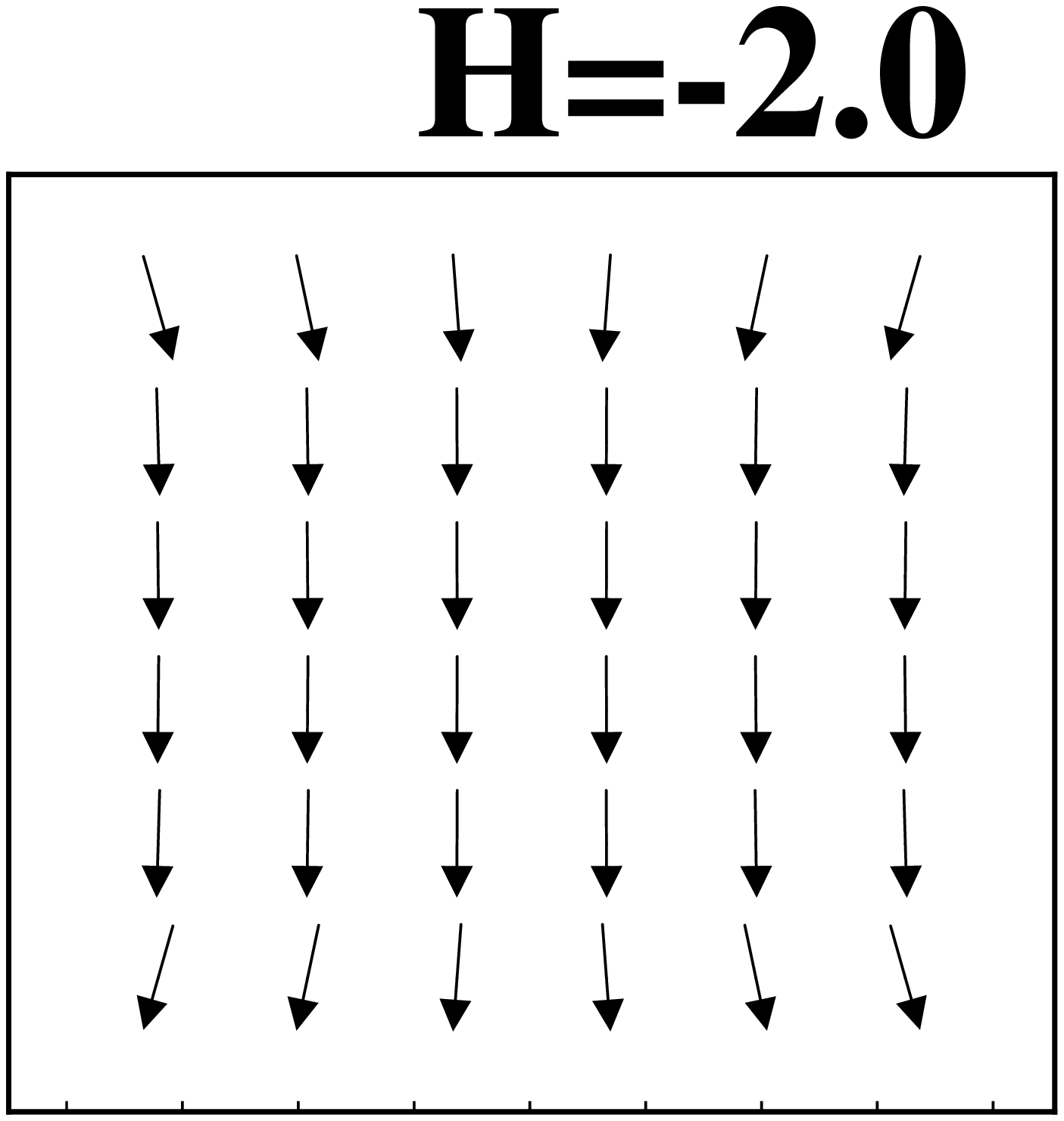}
  \caption{Spin arrangements for an array of $6\times 6$ ferromagnetic 
    nano dots in external magnetic field.}
  \label{fig55}
\end{figure}
For $N=2$, Fig.~\ref{fig2} shows that below $H_0 =M_s$ and for $H_0 \ne 0$ 
the spins form a snake-like domain structure winding clockwise
or counterclockwise. At $H_0 =0$, the array has zero net
per-dot magnetization, due to a vortex-like structure that 
persists for $-3M_s\le H_0\le 0.3M_s$. 

The $N=3$ array was analysed by Camley and
Stamps in \cite{CS1}. This array
also shows snake-like arrangements below
$H_0 =M_s$.  At zero field the final state of the array shows what we call
a ''barrel'' state in which the spins at the left and right columns are
oriented opposite to the central column with a slight tipping of the corner
spins, as shown in Fig.~\ref{fig3}. This agrees with Ref.~\onlinecite{CS1}
except for the tipping of the corner spins. We have made a small angle
expansion of the energy for the spins being nearly aligned, and indeed we
find that the tipped corner spin state is more energy favorable. The
tipping angle was determined by both iteration and analytical
calculations, and agreement is found between the two values. The numerical
value of the corner spin tipping angle, found both by iteration and small
tipping angle analysis, is $|\alpha|=9.1115^\circ$, with numerical error of
order $10^{-5}$. The spin snapshots in Fig.~\ref{fig3} and hysteresis
loop analysis show that the barrel state switches to an inverted barrel
state when the applied field changes sign. 

Fig.\ref{fig4} shows that the $N=4$ array also features snake-like 
arrangements of the spins, for intermediate values of the applied field. 
However, in zero field the total per-dot magnetization is zero, which can 
be attributed to the formation of a vortex in the array's central $2\times2$ 
block. The magnetic moments of the rest of the dots in the array form a ring 
that surrounds the vortex with opposite circulation. This state is stable 
for applied fields $H_0$ satisfying $-0.2M_s \le H_0 \le 0.2M_s$.

For all $N$, the flower state appear at high fields (here, $|H_0|=2.0M_s$). 
The hysteresis loops shown in Fig.~\ref{fig1} show a subtle 
difference in shape between arrays with odd $N$ and arrays  
with even $N$. For odd $N$ the loops show well-defined
jumps whereas for even $N$ this behavior is absent. This
behavior is due to unpaired spins with uncompensated dipole fields.  
The jumps become less apparent as $N$ grows, and will eventually 
disappear for large $N$, where the distinction between even and odd $N$ 
becomes unimportant.

Note that, when an array was placed in zero external field
and given random initial conditions, the solutions converged to the same
states as obtained in the hysteresis-cycle calculations, up to the
degeneracy of the system. Thus, for $N=2, 4$ there are two degenerate
metastable states of opposite chirality (winding) with zero net
magnetization in zero field, each of which has a four-fold 
rotational symmetry. For $N=6$ there are two degenerate states of opposite
chirality, with non-vanishing net magnetization, each of these state has no
apparent rotational symmetry. For $N=3$ there are two degenerate barrel
states, with no rotational symmetry, and $N=5$ is similar to $N=6$. The
hysteresis loop area $A_N$ will produce further evidence that large system
behavior commences with $N=5$ and $N=6$. Metastable states with vanishing
net magnetization may appear for arrays with even $N$. However, our
simulations showed that these states could appear only for $N=2,4$. For
arrays with odd $N$ the unpaired dipoles prevent the formation of such
states.

\section{Hysteresis Loop Area $A_N$ vs. Particle Number $N$}
\begin{figure}[h]
  \centering
  \includegraphics[angle=0,width=3.5in,totalheight=3.0in]{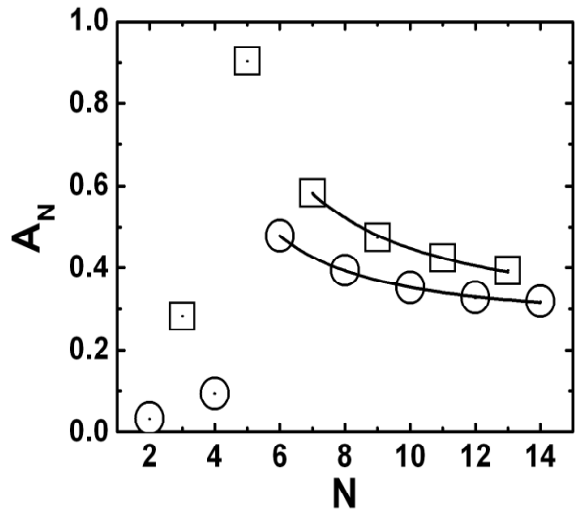}
  \caption{The area of the hysteresis loop as a function of the number of 
    particles $N$.}
  \label{fig6}
\end{figure}

Although the area of the hysteresis loop $A_N$ tends to zero for the
$N=2$ array, it clearly is nonzero for all other arrays.  We present the
hysteresis loop areas in Fig.~\ref{fig6} as circles and squares. From
Fig.~\ref{fig6}, we see that the area of the hysteresis loop decreases
with increasing $N$ except for $N=3$ for $N$ odd and $N=2,4$ for $N$ even.
The $N=5$ and $N=6$ arrays, the first to show something like bulk
behavior, have maximum $A_N$ for odd and even respectively; their spin
arrangements are given in Fig.~\ref{fig5} and Fig.~\ref{fig55}. We have
fitted our data to the asymptotic form 
\begin{eqnarray} 
  A_N=A_\infty +\frac{C}{N^p}
  \label{fit}
\end{eqnarray}
where $A_\infty$, $C$ and $p$ are constants to be determined.  If 
larger values of $N$ had been computationally feasible, we would have 
considered only large values of $N$ for the fit. The 
fit is not as good when $N=5$ is included, so we do not show this case. 
In practice, for $N$ odd the data are fit starting from $N=7$ and for $N$ even
from $N=6$.  Both fits are shown as solid lines in Fig.~\ref{fig6}, where the 
values of $A_\infty$, $C$, and $p$ are given in the table. 

The associated values of $\chi^2$ are both less than $10^{-5}$.
\begin{table}
  \caption{\label{tab:table1}Fitting parameters for data given in
    Fig.~\ref{fig6}, using Eq.~\ref{fit}.}
  \begin{ruledtabular}
    \begin{tabular}{|c|c|c|c|}
      \hline
      N& $A_\infty$ & $C$ & $p$\\
      \hline
      even&$0.278$&$6.31 \pm 0.42$&$1.95 \pm 0.03$\\
      \hline
      odd&$0.278$&$6.92 \pm 0.86$&$1.61 \pm 0.06$\\
      \hline
    \end{tabular}
\end{ruledtabular}
\end{table}
For odd $N$, $A_N$ varies approximately as $A_N \sim N^{-\frac{3}{2}}$ 
whereas for even $N$ it varies approximately as $N^{-2}$.  We attribute 
no fundamental significance to these values. 

\section{Summary}
We have studied the hysteresis and magnetization processes for
$N\times N$ arrays (with $N=2\dots13$) of uniaxial ferromagnetic nano dots  
interacting via the dipole-dipole interaction. For an external 
magnetic field aligned or misaligned with one side of the array, the 
hysteresis loops are surprisingly complex. For arrays with odd $N$ the 
hysteresis loops possess jumps, whereas for even $N$ they do not. As $N$ 
increases, the area $A_N$ of the hysteresis loop begins to saturate, 
approaching a non-zero finite value determined from a data fit. The area of 
the hysteresis loop scales with $N$ approximately as $N^{-\frac{3}{2}}$ for 
$N$ odd, and approximately as $N^{-2}$ for even.

We would like to thank V. L. Pokrovsky, A. S. Kirakosyan and S. Erdin 
for fruitful discussions. This work was supported by NSF grants DMR 
0103455 and DMR 0072115, DOE grant DE-FG03-96ER45598, and Telecommunication 
and Information Task Force at Texas A\&M University.

\end{document}